%% file: main.tex
\documentclass[acmsmall, screen]{acmart}

\input{utils}

\usepackage{hyperref}
\usepackage{amsmath,amssymb,amsfonts}
\usepackage{algorithmic}
\usepackage{textcomp}
\usepackage{multirow}
\usepackage{graphicx} 
\usepackage{subcaption}
\usepackage{booktabs,tabularx,makecell}
\usepackage[table,xcdraw]{xcolor} % To add row colors
\usepackage{array}               % For table customization

\setcopyright{cc}
\setcctype{by}
\acmDOI{10.1145/3808120}
\acmYear{2026}
\acmJournal{PACMSE}
\acmVolume{3}
\acmNumber{FSE}
\acmArticle{FSE113}
\acmMonth{7}
\acmSubmissionID{fse26mainb-p272-p}
\received{2026-02-24}
\received[accepted]{2026-03-24}

\begin{document}

\title{How Do Developers Interact with AI? An Exploratory Study on Modeling Developer Programming Behavior
}

\author{Yinan Wu}
\orcid{0009-0002-1005-2560}
\affiliation{%
  \institution{North Carolina State University}
  \city{Raleigh}
  \country{USA}
}
\email{ywu92@ncsu.edu}

\author{Ze Shi Li}
\orcid{0000-0003-2888-1025}
\affiliation{%
  \institution{University of Oklahoma}
  \city{Norman}
  \country{USA}
}
\email{zeshili@ou.edu}

\author{Kathryn Thomasset Stolee}
\orcid{0000-0003-0584-7094}
\affiliation{%
  \institution{North Carolina State University}
  \city{Raleigh}
  \country{USA}
}
\email{ktstolee@ncsu.edu}

\author{Bowen Xu}
\authornote{Corresponding author.}
\orcid{0000-0002-1006-8493}
\affiliation{%
  \institution{North Carolina State University}
  \city{Raleigh}
  \country{USA}
}
\email{bxu22@ncsu.edu}

\begin{abstract}

% \yinan{how about S[IASE] like a function within 4 parameters instead of S-IASE}\bowen{S-IASE sounds simpler to mention and type?}

% \yinan{cite some papers from FSE}
% \yinan{replication package link update}
Artificial Intelligence (AI) is reshaping how developers adopt software engineering practices, yet the multi-dimensional nature of developer-AI interaction remains under-explored.
Prior studies have primarily examined dimensions  observable from developer activities such as ``Prompt Crafting'' and ``Code Editing,'' overlooking how hidden intentions and emotional dimensions intertwine with concrete actions during AI-assisted programming. 
Understanding the interplay is essential for improving developer experience and future AI assistant designs.
To understand this phenomenon, we conducted a mixed-methods study with 76 developers. We first split developers into AI-assisted and non-AI groups. Each developer performed a programming task (either Python with API management or Java with SQL). Developers retrospectively labeled their self-reported intentions, tool-supported actions, and emotions (on a 7-point valence scale) from screen recordings, supplemented by participant surveys and interviews. Our user study resulted in a novel model, named \textsc{S-IASE}, with four dimensions to describe programming behavior for a given development \textit{state}: \textit{intention}, \textit{action}, \textit{supporting tool}, and \textit{emotion}.
Our analysis reveals several aggregated and sequential behavioral patterns. For example, for aggregated patterns, using AI assistants often led developers to focus more on actively ``creating'' code, evaluating, and verifying the generated results; for sequential patterns, AI-assisted
% participants in comparison to non-AI assisted participants demonstrated greater emotional stability, reflected in more balanced emotional sequences.
participants showed emotionally stable development flows, as opposed to non-AI-assisted participants who experienced more fluctuating emotions.
Interviews revealed further nuance: some developers reported impostor-like feelings, expressing guilt or self-doubt about relying on AI for programming.
The uncovered patterns indicate that our model can provide actionable insights for improving AI assistants' responsiveness, training developers in AI collaboration, and designing developer-centric AI studies. 
% Our contributions include (1) the first multidimensional model of developer-AI behavior, (2) empirical evidence of cognitive-emotional interplay in AI-assisted coding, and (3) an open-source annotation tool for behavioral analysis.
Our work bridges an important gap in understanding the complexities of developer-AI interaction in the programming context and sheds light on future developer-centric research directions.
\end{abstract}

% ``''

% \ccsdesc[500]{Human-centered computing~Empirical studies in HCI}
% \ccsdesc[500]{Software and its engineering~Collaboration in software development}

% \keywords{Developer Programming behavior, Developer-AI Interaction}




% \setcopyright{cc}
% \setcctype{by}
% \acmJournal{PACMSE}
% \acmYear{2026} \acmVolume{3} \acmNumber{FSE} \acmArticle{FSE113}
% \acmMonth{7} \acmDOI{10.1145/3808120}

\maketitle

\section{Introduction}

\input{sections/1-Introduction}

\section{Methodology}
As our goal is to better understand the interactions between developers and AI assistants, we launched a study in which developers performed tasks (with or without AI) and then annotated their hidden intentions in the task recording using our S-IASE model.

\input{sections/3-Experiment-Procedure}

% \section{Data Collection}

\input{sections/3-Data-Collection}

\section{Task Completion Results}\label{sec:taskcompleteratio}

\input{sections/3.5-task-results}

\section{RQ1: Characterizing the Multi-Dimensionality of Developer-AI Interactions}\label{sec:siasemodel}

\input{sections/4-RQ1-taxonomy}

\section{RQ2: Aggregated Behavioral Patterns}

\input{sections/5-RQ2-consistent-results}

\section{RQ3: Sequential Behavioral Patterns}

\input{sections/6-RQ3-pattern}

\input{sections/discussion}

\section{Related Work}

\input{sections/2-Related-Work}

\section{Threats to Validity}

Some threats may impact the validity of our results. We mitigate them as follows: 

\textbf{\emph{Internal validity}} could be impacted because we split participants into two groups based on their prior experience with AI, and not randomly. 
This introduces a potential \textit{selection bias}. 
We also acknowledge the risk of \textit{aptitude bias}, where observed behavioral patterns might be influenced by pre-existing participant traits or a higher baseline aptitude common among developers, rather than being solely dictated by AI usage.
Thus, participants without AI experience may differ from those with AI experience in ways beyond just tool familiarity. 
Furthermore, although stratified sampling helped balance task-related expertise, it may also introduce \textit{residual bias} due to differences in baseline programming skills, prior tool familiarity, or attitudes toward AI between groups. 
While we attempted to mitigate these risks through stratified sampling and post-hoc analysis, we acknowledge that residual confounding factors may remain.
% Moreover, participant trust in AI could change over time. 
We used the pre-survey to help divide the participants into groups, ensuring that participants without or with little AI experience would not be using AI, and those with extensive experience would be using AI.
Similarly, we used participant expertise to determine which activity they were assigned. 
Regarding the potential correlation between AI usage and overall aptitude with the tasks, participants were first explicitly assigned to the AI-assisted or non-AI groups based solely on their self-reported AI usage frequency. After this initial grouping, we employed stratified sampling based on participants' self-reported programming expertise and task familiarity to assign them to one of the two tasks (PyT or JavaT). By doing so, we ensured that each task had participants with varying degrees of familiarity and expertise, thereby reducing the risk of aptitude bias. Nonetheless, we acknowledge that some residual effects from the correlation between AI usage frequency and overall task aptitude could remain as a potential limitation of our study design.

\textbf{\emph{Construct validity}} refers to whether we are asking the right questions in our user study procedure. 
To ensure the construct validity of the study, we carefully designed each stage of our procedure, including the survey questions, the activity tasks, and the interview questions.
The questions and tasks were refined through extensive piloting (see Section~\ref{sec:pilot_study}). 
The self-annotation process relies on participants’ accurate recall and self-assessment. To minimize the impact on the results, we followed Mozannar et al.’s practice~\cite{mozannar2024reading} by instructing participants to annotate their programming behavior right after the session. Participants were also allowed to create labels at their own discretion. 
There may be conceptual and technical differences between the two programming tasks (Java vs. Python). To mitigate this, we collected participants’ task-related prior knowledge via the pre-survey and used it to assign tasks such that both tasks had varied distributions of expertise and familiarity. We acknowledge that Java and Python differ conceptually and technically, but the consistent patterns observed in both tasks suggest these differences did not meaningfully bias our results, so we analyzed them together. Moreover, we chose these tasks as they reflect real-world software development scenarios. Regarding emotion, we used a 7-point valence scale, following the practice of Gardella et al.~\cite{gardella2024performance}; however, we recognize that a 3-level score (Positive, Neutral, Negative) might reduce interpretation subjectivity. 
To mitigate the risk of the tasks that could be directly solved by AI, we designed the tasks that require multi-step human involvement and environment-specific integration and we verified this via our pilot study. Thus, we believe that the threat is minimal. However, we acknowledge that the rapid evolution of LLMs to the solvability of these tasks may change over time.
We acknowledge that the fixed 25-minute task may have introduced time pressure. However, our pilot study determined this time-boxed window to be a practical approximation of real-world development constraints. Our experiments showed that this window was sufficient to capture a rich sequence of interactions for behavioral modeling.

\textbf{\emph{Conclusion validity}} is about the relationship between the treatments and the conclusions. 
Our work explores the potential of our model and identifies important behavioral patterns demonstrated by our participants. 
Our study was conducted with a 25-minute time constraint, which may have amplified differences in action strategies between AI-assisted and non-AI groups. In particular, AI-assisted participants more quickly generated and executed runnable code, whereas non-AI participants may have allocated more time to comprehension and searching. Future work should investigate more action–tool relationships without time constraints. We acknowledge that our validation and member-checking surveys reached a subset of participants, and we have analyzed the minority disagreements. We will explore and expand the explanations in future studies.

\textbf{\emph{External validity}}  may be impacted by participant recruitment: 
we recruited primarily from a university setting, which may limit generalizability. However, 75\% of participants reported prior internship or industry experience, which mitigates this concern to some extent. We acknowledge that this population is better characterized as junior or early-career developers rather than experienced professionals, and thus it is not fully representative of expert developers in industry.
We believe that many of the behaviors we observe are likely to generalize beyond students to early-career developers.
Moreover, we expect our annotation tool to be adaptable for other coding tasks.

\section{Conclusion and Future Work}

In this work, we present a mixed-methods user study with 76 developers, performing programming tasks, to help develop a deeper understanding of developer behavior when interacting with AI assistants. 
Through our user study, we propose a novel model for developer programming behavior that encapsulates four dimensions: \textit{Intention}, \textit{Action}, \textit{Supporting Tool}, and \textit{Emotion}.
Our model provides insights that can be derived from snapshots of the data and uncovered by a sequential view. 
We carefully designed and performed a series of user study experiments, including pilot study, pre-survey, programming activity with self-reporting, post-survey, and interview.
Our findings include both aggregated and sequential behavioral patterns such as AI-assisted participants experiencing tension between productivity and comprehension.

Our study opens up opportunities for future work. Because our tasks emphasized completion within a fixed time window, participants were not expected to produce production-ready code. Thus, our observations capture sequential patterns rather than artifact quality. Future research should investigate how the S-IASE model manifests in settings where code quality, maintainability, and long-term outcomes are central. 
Future research should also investigate these edge cases, such as the annotation ambiguities identified during member checking, to refine the granularity of the S-IASE model and further explore the socio-technical nuances of developer-AI interaction.
Such work could clarify how AI assistance affects not only immediate developer experience but also the robustness and sustainability of software projects.  
With all the valuable findings we uncovered, we advocate that future work should conduct more experiments based on our proposed model to offer practical insights for integrating AI assistants into real-world software engineering processes.

\section*{Data Availability}
The data with personally identifiable information from participants that support the findings of this study are not publicly available due to privacy concerns and restrictions imposed by our Institutional Review Board (IRB).
A replication package containing our annotation tool, pre-survey, post-survey, experiment scripts, and anonymized results is available at our GitHub repository: 
\url{https://github.com/YinanWusoymilk/FSE-2026-How-Developers-Interact-with-AI}.

\section*{Acknowledgments}
This work is funded in part by NSF SHF \#2006947 and \#1749936.

\bibliographystyle{ACM-Reference-Format}
\bibliography{main}

\end{document}

%% file: utils.tex
\usepackage{changepage}  % Adjusts margin indentation
\usepackage{enumitem}

\usepackage{amsmath,amssymb}

\newcommand{\participantquote}[2]{%
\vspace{1mm}
    \begin{adjustwidth}{1.1em}{1.1em} % Smaller indent on both sides
        ``\textit{#1}'' % Italicized quote text
        \textbf{(#2)} % Participant number immediately after
    \end{adjustwidth}
}

\usepackage{url}
\usepackage{tcolorbox}

\newtcolorbox{implicationbox}{
    colback=blue!10,
    colframe=blue!75!black,
    boxrule=1pt,
    left=6pt,
    right=6pt,
    top=3pt,
    bottom=3pt,
    boxsep=0pt,
    arc=2pt, % optional slight rounding
    before skip=4pt, % spacing before box
    after skip=4pt % spacing after box
}

\newtcolorbox{reviewbox}{
    colback=white,       % 背景颜色：白色
    colframe=black,      % 边框颜色：黑色
    fonttitle=\bfseries, % 标题字体：加粗
    boxrule=0.8pt,       % 边框粗细
    sharp corners,       % 直角（不要圆角），模拟 Word 风格
    arc=0mm,             % 圆角半径设为 0
    boxsep=2pt,          % 文字与边框的间距
    left= 4pt,           % 左内边距
    right=4pt,          % 右内边距
    top=4pt,             % 上内边距
    bottom=4pt           % 下内边距
}

%% file: sections/1-Introduction.tex
Software engineering is undergoing a significant shift due to the rise of AI assistants~\cite{kam2025professional, zakharov2025ai, chen2021evaluating}.
AI assistants such as GitHub Copilot~\cite{copilot_users} and ChatGPT~\cite{chatgpt_users} have gained widespread attention and quickly attracted millions of subscribers from students to professional developers alike. GitHub Copilot, in particular, promises to be an ``\textit{AI [code] editor for everyone}''~\cite{github}.
Prior research has explored these AI assistants in various capacities to support software developers throughout the software development lifecycle, ranging from requirements~\cite{terragni2025future, robinson2024requirements},  design~\cite{9653427}, development~\cite{torka2024optimizingaiassistedcodegeneration, 10705649, yeticstiren2023evaluating, 10.1145/3643774}, testing~\cite{10449663}, to even maintenance~\cite{marchezan2024modelbasedmaintenanceevolutiongenai, bidollahkhani2024revolutionizingreliabilityroleai}. 
Due to the wide variety of areas in which a developer can leverage AI, recent large-scale developer surveys found that 76\% of developers in 2024~\cite{stackoverflow_survey} and 84\% of developers in 2025 are already using or plan to use AI in their work~\cite{stackoverflow_survey_2025}.

Yet, as AI assistants become embedded in everyday software programming, a central question arises: \textit{\textbf{how do developers interact with these assistants?}} Without sufficient understanding, we cannot design AI assistants that better align with developer programming behavior and ultimately boost developers' productivity.
Most prior research has focused on ostensible benefits (e.g., reduced repetitive coding ~\cite{ng2024harnessing}) or on narrow contexts (e.g., novice learning~\cite{amoozadeh2024student, Kazemitabaar}). Even when prior studies have looked at interaction, they have often done so from the researchers’ observations, identifying a limited set of modes (e.g., “acceleration” vs. “exploration” modes~\cite{barke2023grounded}). However, few studies have focused on developer-AI interaction from the developer's perspective.
We have limited perspectives on what developers \emph{think} and \emph{feel} when they interact with AI assistants as well as how those perceptions differ from those without AI.

Mozannar et al.~\cite{mozannar2024reading} are among the first to explore developer-AI interaction from the developer's perspective.
They investigated how developers interact with GitHub Copilot during time-restricted coding tasks and developed a taxonomy called CUPS to categorize developer states when interacting with AI assistants. 
% Their insights showed how developer-AI interactions had inefficiencies that could be improved to benefit productivity \cite{mozannar2024reading}. 
While their work represents a significant step toward understanding developer-AI interaction, the CUPS taxonomy treated a ``state'' as a single coding activity and focused on transitions between activities (e.g., from \textit{``Prompt Crafting''} to \textit{``Code Editing''}). 
However, such activity-based labels failed to capture the subtleties and multidimensionality of developer-AI interaction in the context of programming.

By ``\textit{dimension},'' we refer to a distinct, analyzable component of developer-AI interaction (e.g.,~intentions, actions). Each dimension provides a lens to study and compare interactions.
Google's SIA (State–Intention–Action) model~\cite{googlesia} further inspired us to view a programming \textit{state} not as a single coding activity, but as a snapshot of development context, which includes the specific tools, and gives rise to intentions and actions.
These insights led to our understanding that developer-AI programming behavior involves more than one dimension.

This motivated the main question guiding our research: 
\vspace{1mm}
\begin{implicationbox}

How do developers interact with AI assistants in the broader programming context?

\end{implicationbox}
\vspace{1mm}

Our exploratory work aims to understand more deeply how AI assistants shape the different dimensions of developer programming behavior and how these dimensions could be characterized to the sequences of development states.
We conducted a mixed-methods user study with a group of 76 participants with varied backgrounds in terms of education,  programming experience, and familiarity with the topic of the tasks.
The participants were assigned to two groups: AI and non-AI. Participants from the AI group could freely use any AI assistants in their preferred way during task execution. In contrast, those from the non-AI group, who had little to no experience with using AI for programming in the past, performed the task without using AI.
Within each group, participants were assigned to one of two programming tasks: a Java task with MySQL Database Operations or a Python task with GitHub REST API.

From this study, we found that developer programming behavior when interacting with AI involves four distinct dimensions.
We formalize this in our proposed \textbf{S-IASE} model, which characterizes a developer's \textit{\textbf{S}tate}  when they interact with AI assistants, using their \emph{\textbf{I}ntention}, pairs of <\emph{\textbf{A}ction}, \emph{\textbf{S}upporting tool}>, and \emph{\textbf{E}motion}.

\begin{implicationbox}
\centering
state$_{n-1}$ → state$_n$[intention, {<action, supporting tool>}, emotion]  → state$_{n+1}$
\end{implicationbox}

Our model details how intention (e.g., \emph{To Implement Code} $\to$ \emph{To Resolve Errors}), actions supported with tools (e.g., \emph{<Reading and Comprehension, No Tool>} $\to$ \emph{<Crafting Query/Prompt, AI Tool>}), and emotions (e.g., \emph{Neutral} $\to$ \emph{Positive}) manifest throughout developers' states. 
Our proposed model unlocks the new opportunity to perform analysis and reveal behavioral patterns.
Specifically, on top of \textbf{S-IASE}, we performed a mixed-methods analysis and uncovered two complementary types of patterns. 
First, \textit{aggregated patterns} derived from descriptive statistics. For instance, our results show that using AI assistants often makes the programming experience more focused on actively ``creating'' code, evaluating, and verifying the generated results.  
Second, \textit{sequential patterns} identified through automated sequential pattern mining approaches (e.g., CloSpan~\cite{yan2003clospan}). The identified patterns indicate that AI-assisted participants exhibited statistically more stable emotional development flows, whereas non-AI participants experienced more fluctuating emotions.  
Finally, our interview thematic analysis based on socio-technical grounded theory (STGT)~\cite{hoda2024qualitative} revealed more subtle findings: despite this statistical stability, some AI-assisted participants reported impostor-like feelings, expressing guilt or self-doubt about relying on AI. These findings highlight the complex emotional trade-offs in AI-assisted development.

Our work explores the deeper complexities of developer-AI interaction in software development. 
The carefully constructed model and insights drawn from our user study provide valuable guidance for both researchers and practitioners, informing the design of future studies and the development of more effective tools for AI-assisted programming.
The main contributions of our work include:
\begin{itemize}
    \item A model named S-IASE for describing programming behavior, which covers four dimensions: intention, action, supporting tool, and emotion to a given sequential state.
    \item A multi-round user study to iteratively refine our proposed model. Each round included a pre-survey, programming activity, post-survey, and optional interviews.
    \item An Analysis to uncover both aggregated and sequential programming behavior patterns based on our S-IASE model.
    \item An easy-to-use annotation tool to report developer programming behavior.
\end{itemize}

%% file: sections/3-Experiment-Procedure.tex
\subsection{Research Questions}
Our methodology was designed to explicitly address three research questions:

\begin{implicationbox}
\textbf{RQ1}: How can we \textbf{characterize} the multi-dimensionality of developer-AI interactions?
\end{implicationbox}
\noindent To answer this, we developed the S-IASE model that characterizes \textbf{four dimensions} of developer programming behavior: intentions, actions and supporting tools (represented in one or more pairs), and emotions, to the given sequential prior development states. These dimensions were constructed and validated from retrospective self-annotation, surveys, and interviews.  

\begin{implicationbox}
\textbf{RQ2}: What \textbf{aggregated} behavioral patterns emerge for developer-AI interactions?
\end{implicationbox}
\noindent To answer this, we performed a statistical analysis on the distribution of intentions, actions, tools, and emotions across AI-assisted and non-AI groups, identifying consistent differences across tasks.

\begin{implicationbox}
 \textbf{RQ3}: What \textbf{sequential} behavioral patterns emerge for developer-AI interactions?
\end{implicationbox}

\noindent We applied the closed sequential pattern mining algorithm (CloSpan) to the annotated sequences, uncovering recurring sequences for each dimension.  Then we conducted a thematic analysis using STGT~\cite{hoda2024qualitative} on the interviews to capture developers’ perceptions and experiences with AI assistance.

\subsection{S-IASE Model Construction}
To answer the research questions, we developed the S-IASE model as follows. 

\noindent \textbf{Stage 1: Initialization with observations.}
We built a structured model with multiple distinct dimensions, with each dimension representing a perspective along which interaction can be studied and compared.
Our design was inspired by a State–Intent–Action (SIA) model proposed by a Google Research team to characterize software development activities~\cite{googlesia}. It includes but is not limited to three broad categories of developers' activities: (1) debugging and repair tasks, (2) code review tasks, and (3) code editing tasks.
However, the original SIA model is not open-sourced and does not disclose the important details, such as category definitions and whether a user study was conducted internally at Google. 
Moreover, the SIA model is code-centric by design. 
Differently, our study must capture concrete, context-rich behavioral patterns in the programming context from a human-centric perspective to draw conclusions about developer actions and emotions. 

\noindent \textbf{Stage 2: Model Adaptation.}
% Therefore, we make two significant improvements to SIA. 
We developed a new model with additional dimensions and provided clearer definitions of each dimension. 
% Specifically, we first introduce two new dimensions to the model. 
The transition from Google’s SIA model to our proposed S-IASE model was driven by both empirical evidence from our pilot study and established literature.
One new dimension, named \textit{supporting tools}, is paired with \emph{action}.
This dimension was introduced based on the insight that developers in a production environment do not rely on AI exclusively; rather, they conduct actions by navigating a heterogeneous tool ecosystem. We therefore use
supporting tools to distinguish between actions done with or without AI, or any other tools. 
% Our pilot study (described in Section~\ref{sec:pilot_study}) revealed the importance of developer emotions when considering developer-AI interactions.
We also added the \emph{emotions} dimension, informed by prior literature~\cite{cheng2023multi, girardi2020recognizing, graziotin2015feelings, madampe2025emoreflex} and our pilot study (Section~\ref{sec:pilot_study}) to account for affective states that can reflect how emotions and cognitive programming behavior interplay~\cite{girardi2020recognizing} and together shape the multidimensionality of developer-AI interactions.
% Third, we concretize each dimension for further research.
This results in our S–IASE model (\emph{State, Intention, Action, Supporting Tool, Emotion}). 
% Moreover, we also refine our annotation tool based on the feedback from participants recruited for the pilot study.

\noindent \textbf{Stage 3: Dimension labeling.}
To develop the labels within each dimension, we drafted an initial set of labels based on prior studies~\cite{mozannar2024reading, amoozadeh2024student, russell1980circumplex, gardella2024performance}. Then, we allowed participants to add their custom labels during annotation if the predefined options were insufficient.

\noindent \textbf{Stage 4: Validation.}
The model data was collected via the annotation tool (Section~\ref{sec:annotation}) and validated through a pilot study (Section~\ref{sec:pilot_study}) prior to the full study to answer RQ1--RQ3.

\subsection{User Study Procedure}\label{sec:user_study_procedure}

\begin{figure}[t]
    \centering    
    \includegraphics[width=0.8\linewidth]{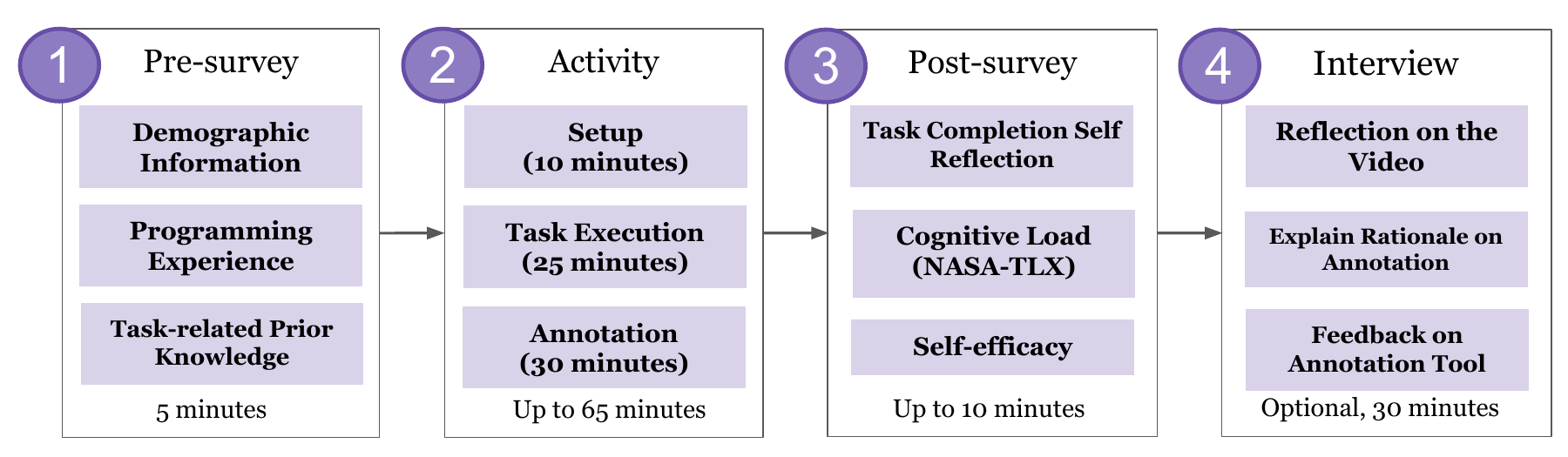}
    \vspace{-3mm}
    \caption{Overview of Our Study Steps}
    \label{fig:experiment_process}
    \vspace{-5mm}
\end{figure}

Figure~\ref{fig:experiment_process} presents the overall steps of our mixed-methods user study with four stages:

\noindent \textbf{S1: Pre-survey.} Each participant completed a pre-survey to obtain their consent to participate, demographic information (e.g., educational background), programming experience (especially on their AI use), and their prior knowledge on the overall topic of our designed tasks (e.g., familiarity with writing and executing SQL queries). We then used the information as a guide to assign participants to different groups (described in Section~\ref{sec:recurtment}).

\noindent \textbf{S2: Activity.}
We provided two options for \textit{Setup} according to participants' preferences: (1)~installing essential tools needed for the activity on their computers locally, such as task-specific programming language interpreters, a necessary database, a screen recording software, and our activity annotation tool. (2)~installing a virtual machine that contains all these essential tools. Participants assigned to the AI-assisted group were explicitly informed that they could freely install and use their preferred AI assistants (e.g., ChatGPT or GitHub Copilot) without restriction, whereas participants in the non-AI group were instructed not to use any AI assistants during the tasks.
Then, we introduced participants to the tasks and group assignment (i.e., AI or non-AI). Participants started  screen recording and then performed their assigned task (\textit{Task Execution}).
After the task was completed, each participant watched their screen recording and used our custom \textit{annotation} tool to break down the whole video into a sequence of intervals. For each interval, participants were asked to label their intentions, actions, supporting tools, and emotions.

\noindent \textbf{S3: Post-survey.} We first asked participants to upload their (1) screen recording of the task execution, (2) annotation results, and (3) implemented code. 
Then, we asked participants to reflect on their experience in completing the task in terms of cognitive load, self-efficacy, and challenges during task execution. 
Lastly, we collected their willingness to participate in the follow-up interview.

\noindent \textbf{S4: Interview (optional).} We invited interested participants to schedule an interview. The interviews helped triangulate our findings from the prior three stages. This additional data collection helped to develop a deeper understanding of participant behavior.

\subsection{Activity Tasks}

We designed two software engineering tasks, \textbf{PyT} and \textbf{JavaT}, based on five criteria.   
First, a task should fit within a 25-minute block. Second, a task should be practical and common in software development. Unlike prior studies that use algorithmic or introductory-level tasks~\cite{mozannar2024reading, zhang2024students}, we took inspiration from Barke et al.'s work~\cite{barke2023grounded} and designed tasks that better reflect real-world software development scenarios. Third, a task should not be directly solved by AI assistants through a single or multiple short prompts without significant developer intervention.
Fourth, a task should require multiple developer activities, such as debugging, bug fixing, and API/database interactions. Fifth, a task should be implemented in popular programming languages.

\noindent$\bullet$ \textbf{PyT: Python Task with GitHub REST API}. This task involves invoking GitHub's REST API in Python to automate organization and repository management over 3 subtasks. 
Participants must create a GitHub organization, use Python to invite a specified owner, and either generate a personal GitHub API token to create 10 public repositories, or automate issue creation in each repository.

\noindent$\bullet$ \textbf{JavaT: Java Task with MySQL Database Operations}. This task involves setting up a MySQL database, implementing Java database operations to interact with the MySQL database, and solving several business queries over 4 subtasks. 

Although the two tasks differ in programming languages and required knowledge, they are comparable in that both represent practical software development activities. Both tasks involve full-stack components and require problem-solving across API use, data processing, and tooling. Importantly, the tasks produce clear, interpretable outputs (e.g., successful API calls, database updates), which help participants immediately see the correctness of their code. This direct feedback supports engagement and motivation during task execution~\cite{horvath2022using}.

%% file: sections/3-Data-Collection.tex
\subsection{Participants and Task Assignment}\label{sec:recurtment}

We recruited participants for our user study in two ways. The first was an in-class mode, where we gave a presentation in both undergraduate- and graduate-level software engineering courses to explain the study's objectives and procedures. 
The second was recruiting participants through the research team's internal connections and snowball sampling. 
To become a participant, one had to be over 18 years old and have prior programming experience.
Our recruitment resulted in 90 participants, with 11 participants recruited via snowballing and 79 from the classroom. 
As compensation, 8 randomly selected participants received \$25 Amazon gift cards.

While AI assistants are becoming ubiquitous for developers, not all developers use them~\cite{stackoverflow_survey, stackoverflow_survey_2025}. 
More importantly, not every developer is \emph{frequently} using AI assistants in their work. 
We recognized the dichotomy between frequent (i.e., more familiarized) and infrequent (i.e., including non-AI) users and separated them into two groups.
Our task assignment intent was to study participants' programming behavior in their more natural environment.

We used stratified sampling to assign each participant to a group based on the following criteria: (1)~If a participant had never or rarely used AI for programming, we assigned them to non-AI groups, i.e., nonAI-PyT and nonAI-JavaT. (2)~Following the practice of Mozannar et al.~\cite{mozannar2024reading}, we assigned participants to a task according to their prior task-related knowledge in the pre-survey. 
We therefore created four participant groups: \textbf{AI-PyT}: participants in this group solved PyT and were allowed to use AI, \textbf{nonAI-PyT}: participants in this group solved PyT and were not allowed to use AI, \textbf{AI-JavaT}: participants in this group solved JavaT and were allowed to use AI, \textbf{nonAI-JavaT}: participants in this group solved JavaT and were not allowed to use AI.
For data filtering, we did not consider a participant's data if the screen recording did not capture the full screen or the annotation log file went missing. Moreover, to prevent the use of secondary devices (e.g., participants in non-AI groups using personal phones to access AI), researchers were present during all sessions. We also excluded four participants who did not complete the tasks during class. Researchers thoroughly examined the videos that participants recorded, and we excluded one participant in the non-AI groups who used AI in experiments.
After filtering, we were left with \textbf{76} participants who completed all the study steps and complied with our instructions. 
Over 76\% of participants reported 3 or more years of programming experience, with a median of over 4 years, and 75\% reported prior internship or industry experience. Full demographic information is available in our replication package~\cite{replication}.

\subsection{Annotation Tool and Procedure}
\label{sec:annotation}
To collect data for the S-IASE model, we developed an annotation tool so participants could divide video recordings into multiple time intervals and annotate each interval based on the model.
Hence, we used the implementation of a tool for annotating programming videos developed by Mozannar et al.~\cite{mozannar2024reading} as our starting point.
We designed the graphical user interface (GUI) and developed features that are essential for our study purpose. 
For instance, following the pilot study, we improved the user experience of the annotation tool by displaying the options with dropdown menus.

\begin{figure}[]
  \centering
  \includegraphics[width=1\linewidth]{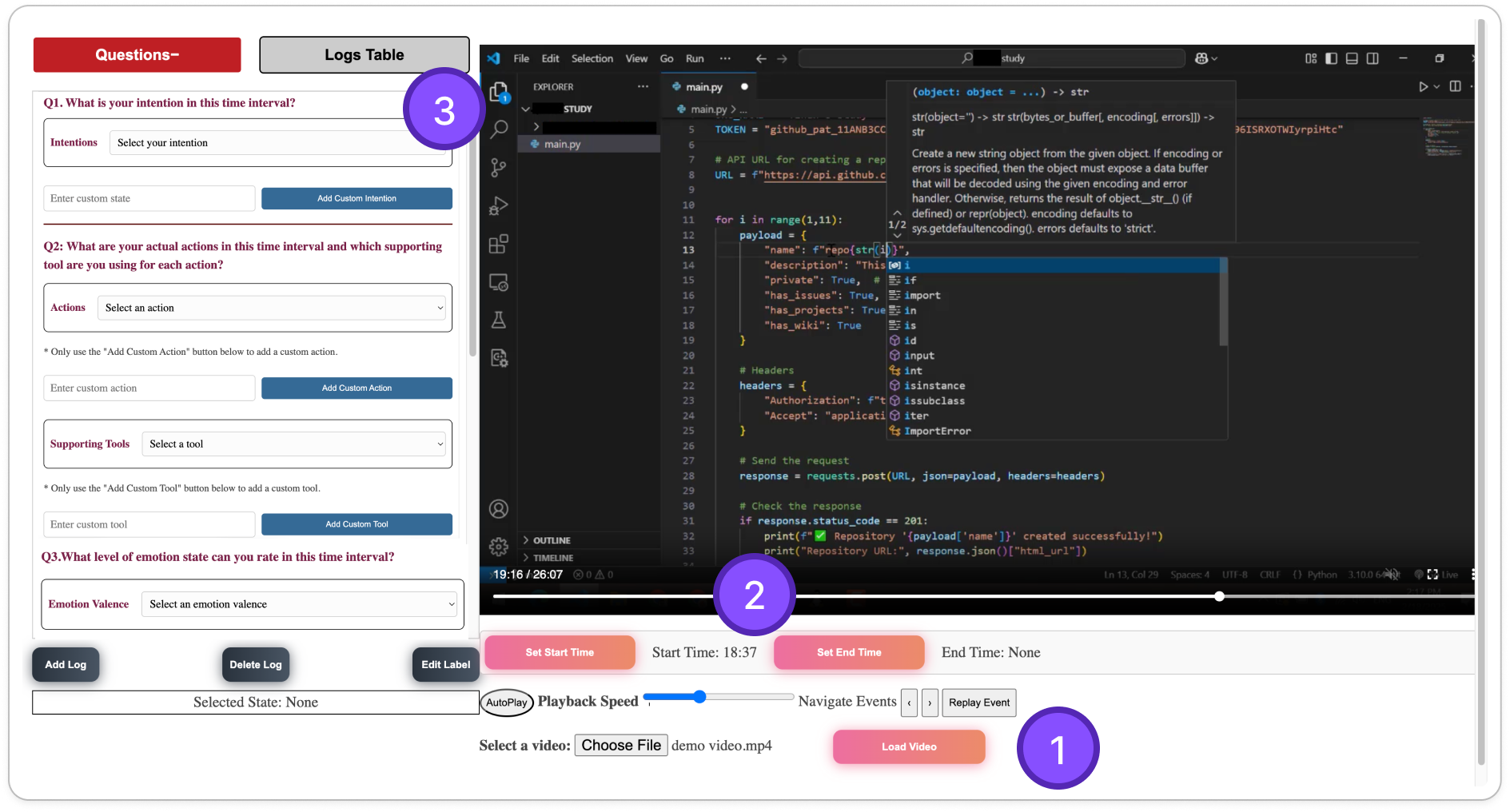} 
  \caption{Our Annotation Tool's GUI: The participant \textbf{(1)} uploads a screen recording of their programming session, \textbf{(2)} sets the start and end time of a specific time interval, and \textbf{(3)} annotates the selected time interval across four dimensions (e.g., intention, action, and emotion) by choosing the corresponding dropdown menus.}
  \label{fig:single_image}
  \vspace{-5mm}
\end{figure}

Figure~\ref{fig:single_image} depicts our annotation tool GUI.
Once the task execution stage was finished, we asked participants to stop recording their screens. 
We then provided a demo of the annotation tool. 
After that, participants annotated their programming session to ensure their recollection of the task remained fresh.
% In Mozannar's original tool, time intervals of the sessions were pre-set. 
Unlike the original tool~\cite{mozannar2024reading}, ours allows participants to define their own time intervals by setting start and end timestamps. 
% We detail our annotation methodology in Section \ref{annotation_methodology}.

In the annotation step, we encouraged participants to annotate approximately 20 intervals (each representing a different time range) describing their development states. This guideline was based on the average number of intervals observed during our pilot study. Although we used 20 intervals as a reference for participants, they could freely create as many intervals as they felt necessary. For each time interval, participants needed to annotate all four dimensions of our S-IASE model (described in Section~\ref{sec:siasemodel}).
In the initial annotation tool, the predefined categories contained \textbf{10 intentions}, \textbf{14 actions}, \textbf{8 supporting tools}, and \textbf{7 emotions} labels. They could also add custom labels within each category if they found predefined options insufficient. 
After data collection, 
two researchers first reviewed a random subset of 10 screen videos and their corresponding annotated labels to assess the completeness and consistency. After validation, the same researchers
conducted a structured merging~\cite{zimmermann2016card} process to consolidate similar custom categories and refine categories based on participant responses. One researcher reviewed and clustered all custom labels, mapping them onto existing categories or creating new categories as necessary (e.g., adding the intention \emph{to set up project environment} and the action \emph{search}). Another researcher independently validated these mappings, and any disagreements (i.e., representing 5\% of all categories) were resolved through discussions until consensus was achieved.
As shown in Table~\ref{tab:taxonomy}, the newly emerged labels are marked with a \textbf{(+)}, indicating that they were added during this merging stage.

\subsection{Pilot Study}\label{sec:pilot_study}

Before experimenting on a larger group of participants, we performed a pilot study with eight participants for two purposes: (1)~design and refine our developer-AI model derived from our literature review (which was \textit{S-IAS} at the time, no \textit{E}), and (2)~refine details in our designed user study steps to ensure clarity for participants.
The pilot study became pivotal for answering \textbf{RQ1} because we discovered through piloting that, in addition to intention and action, \emph{emotion} emerged as a paramount dimension participants repeatedly alluded to.
For example, ``\textit{I am so \textbf{frustrated} that ChatGPT is granting permissions to all the repos via the REST API as opposed to having a more privacy-centric view in its answers.}''
It became clear to us from the pilots that participants demonstrated a fluctuating level of emotions while using AI tools, which is important to consider if we want to improve and strengthen prior characterizations of developer-AI interaction during programming. 
This finding is in line with prior work that has studied the impact of emotion on developer productivity~\cite{madampe2025emoreflex, cheng2023multi, graziotin2015feelings, girardi2020recognizing}. 

As a result, we made a major refinement to our model, adding \textit{emotion} as a dimension. 
This resulted in four dimensions that formed the foundation for our S-IASE model. 
Additionally, piloting helped identify annotation categories for each of these four dimensions. For example, the intention \emph{To Clarify Requirements/Ambiguities} emerged when participants expressed uncertainty regarding task instructions. Actions such as \emph{Logging} and tools like \emph{Database Tools} were added to better capture task-specific activities observed during the piloting sessions.

Piloting also allowed us to refine our task instructions to improve clarity and feasibility. 
For example, the 25-minute task duration was determined based on the performance of pilot participants, who demonstrated that the core requirements of both PyT and JavaT could be addressed within this time-boxed window. For JavaT, we allowed participants to choose between Maven and standard package imports to establish a JDBC connection, reducing setup complexity.
We also improved the annotation tool UI by making it into a single page application with all the annotation categories for a given time interval in a scrollable panel. To prevent data contamination, we confirmed that pilot participants were not included in the main study results.

\subsection{Interview}
As the last stage of our user study, we invited participants to our follow-up interview.
Before the interview, we proposed a set of general interview questions from multiple perspectives. 
% including cognitive load under time constraints, educational implications, and tool usability evaluation. 

For constructing our interview scripts, we mainly focused on \textit{what they thought}, \textit{why they behaved in certain ways}, \textit{how they felt during the experiments}, and \textit{what they had learned}. 
Therefore, the interview scripts were constructed in three layers. 
First, we designed general questions on developer–AI interaction inspired by prior studies (e.g.,~\cite{mozannar2024reading, barke2023grounded, li2024ai, liang2024large, choudhuri2024guides}) and adopted Bloom’s taxonomy~\cite{bloom1956taxonomy, anderson2001taxonomy}
to probe what participants felt they had learned during the task. Second, we focused on cognitive biases and cognitive load under time pressure, using responses from validated instruments such as the NASA-TLX~\cite{hart2006nasa} and self-efficacy scales as anchors for follow-up questions (e.g., “\textit{You rated mental demand as 5, what was happening in that moment?}”). Third, we included questions on the usability of our annotation tool, asking participants to evaluate it and whether it accurately captured their intentions, actions, and emotions. 
Then, for each interview, we carefully watched the participant's uploaded video and read their response history to tailor participant-specific follow-up ``why'' questions, such as \textit{``Why did you switch from using an AI tool to searching Stack Overflow at this point?''} 

Among the 76 participants, 14 (18\%) volunteered to be interviewed.
Each interview took around 30 minutes with two authors and one participant.
Our interviews followed a semi-structured format where we used our prepared interview questions as a general guideline, but allowed flexibility to explore topics that emerged during an interview. 
After obtaining permission from our interviewees, interview audio was recorded and transcribed by Otter.ai~\cite{otterai}.
To ensure that no errors were made in the transcription, one author manually verified the transcript for each interview. 

The same author who conducted each interview and transcription afterwards performed basic data analysis from socio-technical grounded theory (STGT)~\cite{hoda2024qualitative}.
We chose STGT for data analysis due to its grounding in the socio-technical context. 
Part of the reason STGT was developed was due to \emph{``the core SE practice of programming is not strictly technical, rather, it is socio-technical, involving intensive human-human collaboration and coordination and human-technology interactions.''}~\cite{hoda2021socio}
With the addition of AI assistance to programming, the subject matter we studied is fittingly in the socio-technical domain. 
In contrast to traditional grounded theory methods, which are dedicated to theory development and do not support limited applications for data analysis, STGT can support data analysis to generate rich descriptive themes and categories.
These themes and categories helped supplement our proposed developer-AI model and findings about observed developer behavior. % \yinan{i think we can adopt several validation strategies, one member checking, or conduct few more interviews.}
Furthermore, we adopted a reflexive, researcher-led approach to conduct STGT from an interpretivist perspective~\cite{hoda2024qualitative}. 
The first author served as the primary analytic instrument, conducting open coding, constant comparison, and iterative memoing across all interview transcripts. This process resulted in a range of themes. For this paper, we present themes most aligned with our research focus, such as \textit{``Trust but Verify.''} Member checking with a subset of 6 interviewees helped confirm the resonance and validity of the presented interpretations.

%% file: sections/3.5-task-results.tex
\begin{table}[t]
    \centering
    \caption{Task Completion Results (N = 76)}
    \vspace{-2mm}
    \tiny
    \label{tab:task_completion}
    \scriptsize
    \setlength{\tabcolsep}{1.2pt} 
    \renewcommand{\arraystretch}{0.8}

    % -------- 第一排：Python Task --------
    \resizebox{\columnwidth}{!}{%
    \begin{tabular}{lcccccccc}
    \toprule
    \multicolumn{9}{c}{\textbf{Python Task (3 Subtasks, N = 39)}} \\
    \midrule
     & \multicolumn{2}{c}{Step 1} & \multicolumn{2}{c}{Step 2} & \multicolumn{2}{c}{Step 3} & \multicolumn{2}{c}{All Subtasks} \\
    \cmidrule(lr){2-3} \cmidrule(lr){4-5} \cmidrule(lr){6-7} \cmidrule(lr){8-9}
     & Completed & Correct & Completed & Correct & Completed & Correct & Completed & Correct \\
    \midrule
    AI-Assisted & 31/31 & 15/31 & 27/31 & 21/31 & 22/31 & 16/31 & 22/31 & 8/31 (26\%) \\
    Non-AI      & 8/8   & 1/8   & 4/8   & 1/8   & 0/8   & 0/8   & 0/8   & 0/8 (0\%) \\
    \midrule
    Overall     & 39/39 (100\%) & 16/39 (41\%) & 31/39 (80\%) & 22/39 (56\%) & 22/39 (56\%) & 16/39 (41\%) & 22/39 (56\%) & 8/39 (21\%) \\
    \bottomrule
    \end{tabular}
    }

    \vspace{4mm} % two tables differences

    % -------- 第二排：Java Task --------
    \resizebox{\columnwidth}{!}{%
    \begin{tabular}{lcccccccccc}
    \toprule
    \multicolumn{11}{c}{\textbf{Java Task (4 Subtasks, N = 37)}} \\
    \midrule
     & \multicolumn{2}{c}{Step 1} & \multicolumn{2}{c}{Step 2} & \multicolumn{2}{c}{Step 3} & \multicolumn{2}{c}{Step 4} & \multicolumn{2}{c}{All Subtasks} \\
    \cmidrule(lr){2-3} \cmidrule(lr){4-5} \cmidrule(lr){6-7} \cmidrule(lr){8-9} \cmidrule(lr){10-11}
     & Completed & Correct & Completed & Correct & Completed & Correct & Completed & Correct & Completed & Correct \\
    \midrule
    AI-Assisted & 31/31 & 24/31 & 25/31 & 8/31 & 13/31 & 5/31 & 10/31 & 2/31 & 10/31 & 2/31 (7\%) \\
    Non-AI      & 6/6   & 1/6   & 2/6   & 0/6   & 0/6   & 0/6   & 0/6   & 0/6   & 0/6   & 0/6 (0\%) \\
    \midrule
    Overall     & 37/37 (100\%) & 25/37 (68\%) & 27/37 (73\%) & 8/37 (22\%) & 13/37 (35\%) & 5/37 (14\%) & 10/37 (27\%) & 2/37 (5\%) & 10/37 (27\%) & 2/37 (5\%) \\
    \bottomrule
    \end{tabular}
    }
\end{table}

Table~\ref{tab:task_completion} details how our participants performed on each of the subtasks for PyT and JavaT. 
We distinguished between Completed (the participant performed the required actions and finished the specific subtask) and Correct (the outcome of the subtask was verified to be functionally accurate).
A participant achieved ``All Subtasks Correct'' if they completed and correctly solved all subtasks within 25 minutes.
Among 39 participants for PyT, only 8 were successful in solving all subtasks, while only 2 out of 37 were successful for JavaT.
All 10 participants who correctly solved all their assigned subtasks were from AI-assisted groups.
We found that 42\% (32/76) of all participants successfully reached the final subtask of their assigned activity (56\% for PyT and 27\% for JavaT). However, the overall rate for full functional correctness remained at 13\% (10/76). This suggests that while the tasks were possible to navigate within the time limit, the complexity of API and database interactions presented a substantial challenge to functional accuracy. The lower success rate in the Java task, compared to the Python task, further indicates that environment setup and multi-step database operations imposed a heavier correctness burden on developers.
Moreover, we found that the AI group achieved a higher completion rate than the non-AI group for each subtask.
The results also indicate that while AI markedly boosted both completion and correctness, the complexity of the tasks presented a substantial challenge within the 25-minute limit.

On average, participants in AI-assisted groups adopted more than one AI assistant (1.37) during the experiment. ChatGPT was by far the most popular, used by 38 participants, with an additional 13 participants relying on the paid version. 12 participants used GitHub Copilot as their AI assistant in IDE. This distribution indicates multiple assistants' strategies for adopting AI for task completion.

\begin{table}[tb]
\centering
\vspace{-2mm}
\caption{NASA-TLX Scores for AI-Assisted and Non-AI Groups}
\vspace{-3mm}
\label{tab:nasa-split}
\scriptsize                    
\setlength{\tabcolsep}{1pt}        % 压缩列间距
\renewcommand{\arraystretch}{0.95} % 压缩行高

% 左列固定宽度以触发自动换行；数字列用窄 p 列居中
\begin{tabularx}{\linewidth}{>{\raggedright\arraybackslash\bfseries}p{0.45\linewidth}
                                  *{3}{>{\centering\arraybackslash}p{0.085\linewidth}}
                                  *{3}{>{\centering\arraybackslash}p{0.085\linewidth}}}
\toprule
\makecell[l]{\textbf{NASA-TLX Items}} &
\multicolumn{3}{c}{\textbf{AI-Assisted (N=62)}} &
\multicolumn{3}{c}{\textbf{Non-AI (N=14)}} \\
\cmidrule(lr){2-4} \cmidrule(lr){5-7}
 & Mean & Std & Med. & Mean & Std & Med. \\
\midrule
Mental Demand: How mentally demanding was the task? & 3.90 & 1.45 & 4 & 3.64 & 1.28 & 3 \\
Physical Demand: How physically demanding was the task? & 2.24 & 1.49 & 2 & 1.57 & 1.09 & 1 \\
Temporal Demand: How hurried or rushed was the pace of the task? & 4.98 & 1.74 & 5 & 5.64 & 1.08 & 6 \\
Performance: How successful were you in accomplishing what you were asked to do? & 4.11 & 2.19 & 4 & 1.86 & 1.66 & 1 \\
Effort: How hard did you have to work to accomplish your level of performance? & 3.63 & 1.53 & 4 & 4.14 & 1.29 & 4 \\
Frustration: How insecure, discouraged, irritated, stressed, and annoyed were you? & 3.82 & 1.84 & 4 & 5.21 & 1.12 & 5 \\
\bottomrule
\end{tabularx}
\vspace{-3mm}
\end{table}

We also present the results of cognitive load using NASA-TLX scores~\cite{hart2006nasa} from the post-survey (Table~\ref{tab:nasa-split}). 
Participants who used AI reported higher mean scores for perceived performance, while those in the non-AI group reported higher temporal demand, effort, and frustration. 
AI-assisted participants also rated physical and mental demand slightly higher than non-AI participants. 
As for self-efficacy, participants in AI-assisted groups reported a mean score of 3.06 for satisfaction with the quality of their solutions, and a mean score of 3.34 for confidence in solving similar coding problems. In contrast, participants in non-AI groups reported a lower mean score of 1.57 for satisfaction and a mean score of 2.07 for confidence.

%% file: sections/4-RQ1-taxonomy.tex
\subsection{Outcome}

Our participants were able to add labels to the model as needed. After data collection and structured merging, we indentified 12 intentions, 16 actions, 8 supporting tools, and 7 emotions. Newly emerged labels are marked with \textbf{(+)} to indicate they were added during the merging stage, as shown in Table~\ref{tab:taxonomy}.

\textbf{\emph{State}}
 represents the current development context, situated within the task and environment. Unlike prior taxonomies~\cite{mozannar2024reading, amoozadeh2024student} that predetermine states or general activities (e.g., ``Waiting For Suggestion'' or ``Prompt Crafting''), we infer states dynamically from a participant's codebase, observed actions, and tool usage~\cite{googlesia}.
We argue that the \emph{State} dimension serves as the foundation for developer behavior:
(1) Developers form an \emph{intention} based on the current state of the environment. For instance, if the terminal displays a compilation error, the developer may form the \emph{intention} to resolve the error.
(2) The developer then translates this \emph{intention} into an \emph{action}, such as editing code or searching for a solution using different \emph{supporting tools}, while experiencing a specific \emph{emotion}.
(3) This combination of \emph{intention, action, and emotion} leads to a transition to a new state (e.g., implementing a new function or encountering a new bug), creating a feedback loop where the updated state influences the next intention and action.

As developers often switch between subtasks (e.g., writing a new snippet and then running tests to debug), states can overlap~\cite{mozannar2024reading}. We do \emph{not} provide a formal state model here; we argue that the state dimension is inherently fluid and context-dependent. 
However, recognizing the \emph{State} as a dynamic input is crucial for understanding how developers form intentions, take action within assisted tools, and experience emotional responses during software development.

\noindent$\bullet$ \textbf{Intention}
reflects a developer’s goal for software development~\cite{yu2025test, alomar2024refactor, googlesia}. We argue that the \emph{Intention} is the motivation (e.g., the goal is to do something) behind every one of their physical or digital actions. 
Hence, \emph{Intentions} are often invisible, while actions are usually visible. For example, a developer intends to understand the task description when given a task, but the developer's action could vary from prompting AI tools to understanding or using a search engine to retrieve relevant information on the Internet. 
% We identified 12 intention categories through iterative coding of developer-AI interactions. 
Drawing on our pilot study and participant task activity annotations, 
% researchers' domain expertise, 
we identified \emph{12} \emph{Intentions}, as shown in Table~\ref{tab:taxonomy}.

\noindent$\bullet$ \textbf{Action} describes how a developer proceeds toward a specific intention. 
We identified 16 \emph{Action} categories, as shown in Table \ref{tab:taxonomy}.
Some of these (e.g., \emph{Writing New Code}, \emph{Running Tests}) are readily visible 
in screen recordings, while others (e.g., \emph{Reading and Comprehension}) are partially inferred from user behavior (e.g., scrolling or paused reading) and participant self-annotations.

\noindent$\bullet$ \textbf{Supporting tools}
indicate whether and how a developer uses external tools during an action. 
Prior studies show that actions differ significantly when AI is involved. 
To capture this distinction, we separate \emph{Action} from \emph{Supporting Tools}.
Given the nature of our study, AI assistants were frequently relied upon to support an \emph{Action}.

\noindent$\bullet$ \textbf{Emotion} is measured by a 7-point valence scale as shown in Table~\ref{tab:taxonomy}.
While it is hard to capture instant feelings without external tools such as eye-tracking devices or brain cognitive analysis tools, we adopted Russell's circumplex model of affect~\cite{russell1980circumplex} to measure emotions from an exploratory perspective, and we followed the practice of Gardella et al.~\cite{gardella2024performance} to adopt a 7-point valence scale.
% We therefore extend the SIA model with Emotion to 
We prioritized valence due to its practicality in retrospective annotation~\cite{gardella2024performance}. 
For example, a developer might report a negative valence after repeated failed attempts to resolve errors with AI assistants.

\subsection{Validation and Member Checking}

To validate our proposed model category set, we employed two validation approaches.
First, in the post-survey, we included an optional validation question on a 5-point Likert scale: ``The multiple input fields (e.g., three input boxes) capture my thoughts (e.g., intentions and emotions) and actions effectively.'' 
Among the 32 participants who responded, 4 (12\%) strongly agreed, 19 (59\%) agreed, 5 (16\%) were neutral, 3 (9\%) disagreed, and 1 (3\%) strongly disagreed, indicating that most participants considered the input fields reflective of their actual programming behavior.
We further conducted member checking~\cite{harper2012member, ford2017characterizing} with the final category set. Among the 20 participants who responded, 15 strongly agreed with our merged labels, while the remaining participants expressed uncertainty. Through follow-up inquiries, we found that these 5 participants did not strongly disagree or disagree with the existing categories but rather struggled with the inherent ambiguity of their own latent intentions in a few programming states. They noted that while our labels were the most appropriate available, the ``hidden'' nature of programming intentions occasionally made precise self-annotation difficult. This underscores that while S-IASE provides a structured model, capturing the full spectrum of developer programming behavior, especially intricate intentions, remains an exploratory challenge.

\input{tables/taxonomy}

%% file: tables/taxonomy.tex
\begin{table*}[t]
\caption{Proposed S-IASE Model}
\label{tab:taxonomy}
% \vspace{-2mm}
\tiny
\renewcommand{\arraystretch}{0.9}
\setlength{\tabcolsep}{4pt}
\begin{tabular}{p{0.26\linewidth} p{0.22\linewidth} p{0.21\linewidth} p{0.17\linewidth}}
\toprule
\multicolumn{4}{c}{\textbf{State (with four dimensions)}} \\ 
\midrule
\textbf{D1: Intention} & \textbf{D2: Action} & \textbf{D3: Supporting Tools} & \textbf{D4: Emotion} \\
\midrule
I1 To Understand Context          & A1 Configuration (+)                       & T1 Database Tools                & E1 Extremely Positive \\
I2 To Design/Plan New Code        & A2 Reading and Comprehension           & T2 Search Engine                 & E2 Positive \\
I3 To Implement Code              & A3 Executing the Code                  & T3 AI Tools                      & E3 Slightly Positive \\
I4 To Set Up Project Environment (+) & A4 Waiting                             & T4 Proprietary Dev. Platform     & E4 Neutral \\
I5 To Clarify Requirements/ Ambiguity & A5 Crafting Query/Prompt               & T5 Compiler                      & E5 Slightly Negative \\
I6 To Explore Alternative Solutions & A6 Writing New Code                  & T6 Static Program Analysis       & E6 Negative \\
I7 To Resolve Errors              & A7 Editing Existing Code               & T7 IDE                           & E7 Extremely Negative \\
I8 To Verify Existing Code        & A8 Writing Documentation               & T8 Not using any tools           & \\
I9 To Evaluate Suggestions        & A9 Watching Program Status             &                                   & \\
I10 To Optimize Code              & A10 Accepting results from tools       &                                   & \\
I11 To Understand Required Knowledge (+)& A11 Refining results from tools      &                                   & \\
I12 Do Nothing                    & A12 Rejecting results from tools       &                                   & \\
                                  & A13 Logging                            &                                   & \\
                                  & A14 Editing Query/Prompt               &                                   & \\
                                  & A15 Running Tests                      &                                   & \\
                                  & A16 Search (+)                             &                                   & \\
\bottomrule
\end{tabular}
\vspace{-2mm}
\end{table*}

%% file: sections/5-RQ2-consistent-results.tex
\begin{figure*}
    \centering
    \includegraphics[width=\linewidth]{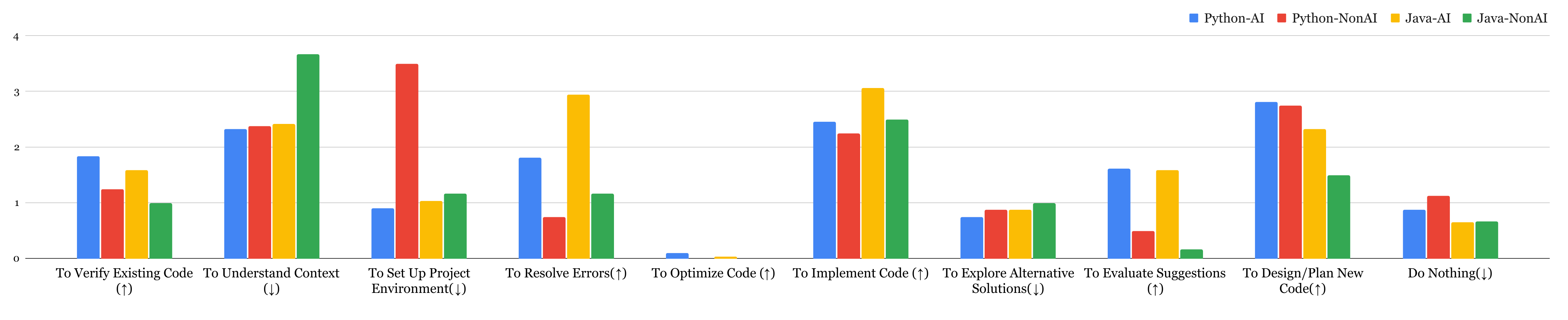}
    \vspace{-6mm}

    \caption{Mean number of intention occurrences per participant across the four groups. The x-axis shows intention categories and the y-axis shows mean counts per participant. (↑) indicates more frequent in AI-assisted group than non-AI group; (↓) indicates less frequent in AI-assisted group than non-AI assisted group.}

    \label{fig:intent-res}
    \vspace{-1mm}
\end{figure*}
\begin{figure*}
    \centering
    \includegraphics[width=\linewidth]{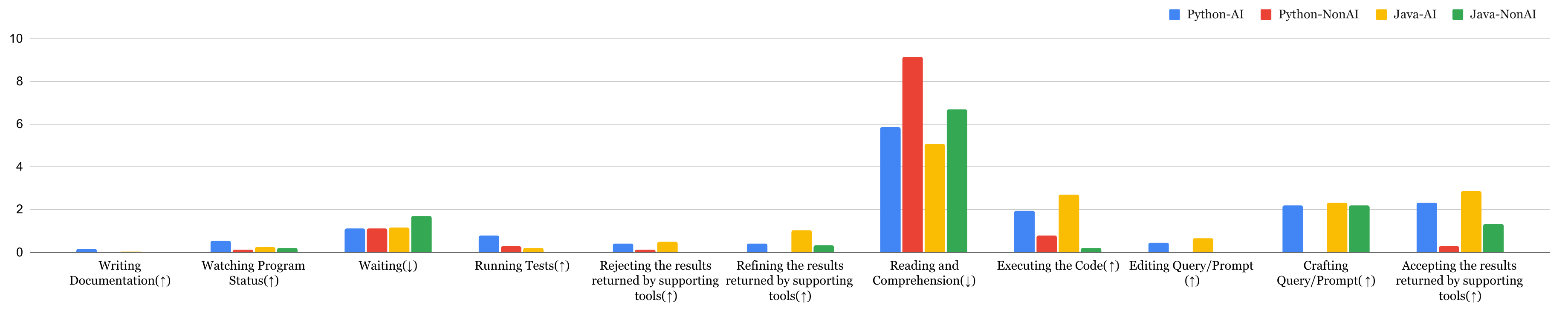}
    \vspace{-6mm}
    \caption{Mean number of action occurrences per participant across the four groups. The x-axis shows action categories and the y-axis shows mean counts per participant. (↑) indicates more frequent in AI-assisted group than non-AI group; (↓) indicates less  frequent in AI-assisted group than non-AI assisted group.}

    \label{fig:action-res}
    \vspace{-1mm}
\end{figure*}
\begin{figure}[]
\centering
\begin{subfigure}{0.49\linewidth}
  \centering
  \includegraphics[width=\linewidth]{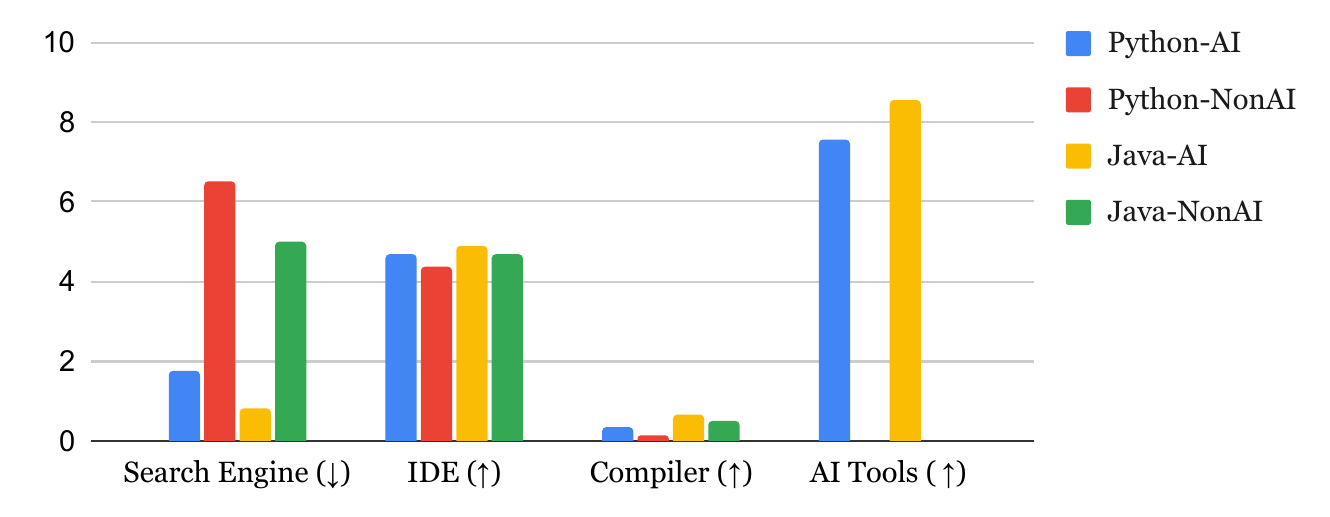}
  \caption{Supporting Tool}
  \label{fig:tool-res}
\end{subfigure}\hfill
\begin{subfigure}{0.49\linewidth}
  \centering
  \includegraphics[width=\linewidth]{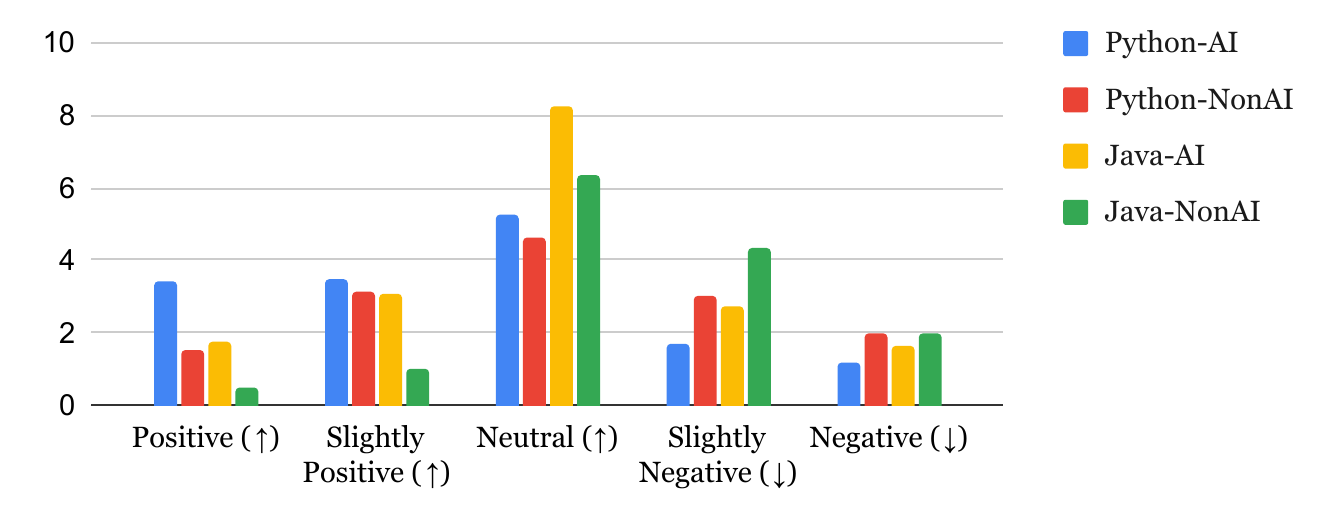}
  \caption{Emotion}
  \label{fig:emo-res}
\end{subfigure}
    \caption{Mean supporting tool and emotion occurrences per participant across the four groups. The x-axis shows categories and the y-axis shows mean counts per participant. (↑) indicates more frequent in AI-assisted group than non-AI group; (↓) indicates less  frequent in AI-assisted group than non-AI assisted group.}
\label{fig:tool-emo-combo}
\end{figure}

To address RQ2, our analysis focused on identifying the differences between AI-assisted and non-AI participants.
To make the interpretation easier, we visualize our results with explanations.
% We provide all the detailed numbers in our replication package.
The bar charts in Figure~\ref{fig:intent-res}, Figure~\ref{fig:action-res}, and Figure~\ref{fig:tool-emo-combo} present the respective results for the aggregated behavioral patterns that emerged across \textit{intention}, \textit{action}, \textit{supporting tool}, and \textit{emotion} dimensions. 
For each category, we first report its occurrence frequency by collecting the number of occurrences and calculating the mean number of occurrences for each participant. We computed the relative difference per task and averaged them. A positive value means the category is more (↑) frequent in the AI groups, and a negative value means less (↓).
Besides, we discuss patterns with statistically significant differences ($\alpha = 0.05$). Specifically, for the categorical dimensions, we computed per-participant rates by dividing the number of intervals annotated with a given category by the total intervals per participant. For emotion, we calculated each participant’s mean valence score. We then fitted ordinary least squares (OLS) regression~\cite{cohen2013applied} with heteroscedasticity-robust standard errors (HC3)~\cite{long2000using}. All models controlled for AI condition and task type. For the categorical dimensions, we applied Benjamini–Hochberg false discovery rate (FDR) correction~\cite{benjamini1995controlling} to account for multiple comparisons, and primarily discuss categories that remained statistically significant after correction.

\noindent$\bullet$ \textbf{Intention}. We found that 6 of the 12 intentions occurred more frequently among AI participants than among non-AI participants. They are: \textit{To Verify Existing Code}, \textit{To Resolve Errors}, \textit{To Optimize Code}, \textit{To Implement Code}, \textit{To Evaluate Suggestions}, and \textit{To Design/Plan New Code}. 
Specifically, AI-assisted participants formed the intention  \textbf{\textit{To Evaluate Suggestions}} significantly more frequently ($\beta = 0.069, p < 0.001 $). This suggests that using AI assistants often leads to occurrences of participants seeking to assess and evaluate their code. Other intentions, such as \textbf{\textit{To Resolve Errors, To Implement Code, and To Design/Plan New Code}}, showed directional differences between AI-assisted and non-AI groups but did not reach statistical significance after correction. These patterns therefore suggest descriptive trends that programming with AI keeps participants focusing on actively ``creating'' code (i.e., with AI assistance), such as designing, evaluating, and resolving the generated results (e.g.,  \textbf{\textit{To Resolve Errors}} ($\beta = 0.0748$)).

\noindent$\bullet$ \textbf{Action}. Among the 16 actions, we found that 9 of them occurred more frequently with AI-assisted participants. After statistical testing and multiple-comparison correction, a subset of actions exhibited statistically significant differences. Particularly, \textbf{\textit{Executing the Code}} ($\beta = 0.096, p < 0.001$) and \textbf{\textit{Editing Query/Prompt}} ($\beta = 0.028, p < 0.001$) occurred significantly more frequently. This can be explained by the identified vibe coding phenomenon~\cite{vibecoding, sarkar2025vibe}. Interestingly, with this phenomenon, we found that they performed \textbf{\textit{Reading and Comprehension}} significantly less frequently ($\beta = -0.146, p < 0.001$). These results suggest that AI assistants transform participants’ programming experience into an iterative cycle of prompting and rapid verification (trial-and-error), reducing the time spent on deep manual code reading and environment setup.
For example, when AI-assisted participants implemented PyT on querying  GitHub's API service, they often directly copy-pasted the AI-generated code and then performed quick verification by executing the code without taking specific actions to try and understand the code. 
If the code functioned as expected, they directly progressed to the next step of the task. 
\textbf{This pattern suggests that participants often perceived AI outputs as sufficiently usable to proceed without deep inspection}. 
This result is also consistent with our findings in terms of task completion rates, i.e., the AI-assisted group achieved a higher completion rate than the non-AI group (described in Section~\ref{sec:taskcompleteratio}). 
 
\noindent$\bullet$ \textbf{Supporting Tool}. We observed that AI-assisted developers relied significantly less on Search Engines ($\beta = -0.246, p < 0.001$). This is in line with the decline in internet traffic to Stack Overflow since the release of ChatGPT~\cite{stackoverflowtraffic}, and suggests that AI assistants have become the primary knowledge interface. Together with the action-level results, this pattern indicates that AI assistance promotes an iterative “prompt–run–check” practice, reducing time spent on deep manual reading and external information seeking. Under this practice, tool use is primarily oriented toward rapid execution and verification rather than extended web searching.

\noindent$\bullet$ \textbf{Emotion}. We found that \textit{Positive}, \textit{Slightly Positive}, and \textit{Neutral} emotions were more commonly expressed by AI-assisted participants, while \textit{Negative} and \textit{Slightly Negative} emotions were more common for non-AI-assisted participants. 
Primary OLS results showed that AI assistance was associated with significantly more positive reported emotions. The AI group reported significantly more positive emotions ($\beta = 0.632, p = 0.016$) compared to the non-AI group. This indicates that participants allowed to use AI assistants for the programming task \textbf{experienced more positive feelings than those without AI}.

To assess the robustness of our findings, we conducted an interval-level analysis using mixed-effects models (LMM)~\cite{laird1982random} for emotion and generalized estimating equations (GEE)~\cite{liang1986longitudinal} for categorical dimensions, accounting for participant-level clustering. Across the Intention dimension, the same core categories remained significant with consistent effect directions. For Action, all categories significant in the robustness check were also identified in the primary analysis, while the primary analysis identified a small number of additional action categories. For Supporting Tools, the robustness check identified \textit{Proprietary Developer Platform} as an additional significant category. Importantly, no statistically significant findings in the primary analysis were contradicted by the robustness checks.
All detailed results are provided in our replication package~\cite{replication}.

%% file: sections/6-RQ3-pattern.tex
To answer RQ3, we conducted a \textbf{sequential pattern mining} analysis 
with a three-step, mixed-methods procedure: 
\textbf{(1) Automated pattern mining:} For each S-IASE dimension, we applied CloSpan, a closed sequential pattern mining algorithm, based on the annotated AI-assisted intervals produced by 62 of the 76 participants. We chose CloSpan because it enumerates closed patterns via a prefix-projected depth-first search strategy, producing a non-redundant set of frequent sequences~\cite{yan2003clospan}. Prior studies have applied closed (or frequent) sequential pattern mining to developer data (e.g., IDE usage mining~\cite{damevski2016mining} and API usage~\cite{xie2006mapo}), demonstrating its suitability for analyzing developer behavioral data. We applied the CloSpan algorithm with a minimum support threshold of 10\%, in line with prior work~\cite{damevski2016mining}, to the four dimensions of our S-IASE model: Intention, Action and Supporting Tool Pair(s), and Emotion. This step provided a non-redundant set of frequent candidate sequential patterns.
\textbf{(2) Manual interpretation and selection of patterns}: Two researchers independently inspected the frequent patterns produced by Step 1. For each pattern, they: (i) examined its occurrences in the underlying participant sequences and screen-recording context, (ii) assessed whether the pattern represented a coherent, interpretable behavioral regularity (e.g., the “Trust but Verify” sequence To Understand Context → To Implement Code → To Design/Plan New Code), and (iii) grouped semantically similar patterns. Disagreements about inclusion or grouping were resolved through discussion until consensus was reached. This step reduced the raw mined output to a smaller set of behaviorally meaningful patterns that we considered for reporting.
\textbf{(3) Triangulation with interview data and existing knowledge}: We then triangulated these candidate patterns with our 14 follow-up interviews (analyzed using STGT) and existing concepts in prior literature (e.g., vibe coding~\cite{sarkar2025vibe, vibecoding}, trust in AI assistants~\cite{choudhuri2024guides}, and impostor phenomenon~\cite{guenes2024impostor}). Patterns were retained only when supported by both sequential data and qualitative evidence. Through this process, the sequential patterns are grounded in both quantitative evidence from the sequence mining algorithm and qualitative evidence from interviews and prior knowledge.

\subsection{Trust but Verify: Navigating Confidence and Skepticism in AI-Assisted Coding}

Our analysis revealed that developer interactions with AI assistants often follow a ``trust but verify'' sequence, where developers engage with AI-generated suggestions but maintain a critical stance. 
\begin{small}  
\participantquote{\textbf{I like using the body part of the code that ChatGPT generated} — it's usually simple and easy to understand. Then I want to write the rest myself because it’s \textbf{easy for me to verify.}}{P2}
\end{small}
We observed a complex relationship between task familiarity, trust in AI, and the need for human verification.
This pattern can be characterized as:

\begin{implicationbox}
    \emph{To Understand Context} → \emph{To Implement Code} → \emph{To Design/Plan New Code}
\end{implicationbox}

\noindent This pattern is particularly noteworthy because it reverses the traditional ``Design-then-Implement'' paradigm. With AI assistance, developers often leverage AI to implement code first to gain a concrete foothold in the codebase, and then proceed to refine the architecture or design (i.e., \textbf{\textit{To Design/Plan New Code}}) based on the AI's output. We also found that the willingness to adopt an AI assistant is influenced by task familiarity. 
Prior work~\cite{barke2023grounded} indicates that people are more willing to adopt AI assistants for familiar tasks; we extend this by finding that developers were more likely to accept AI-generated suggestions when they perceived a task as either highly familiar or highly unfamiliar.
For tasks that participants were familiar with but were not fluent in developing (e.g., setting up and establishing a MySQL Database connection), participants exhibited skepticism toward AI assistants despite correct suggestions. 
They often hesitated to trust AI-generated suggestions without external validation.
% In particular, participants (P6, P9) engaged in defensive verification tactics for these situations that were dismissive towards AI suggestions. 
% For example, when P6 prompted ChatGPT for setting up a REST API token, ChatGPT provided the correct link to the relevant GitHub page.
For example, P6 did not immediately trust the link ChatGPT generated and instead Googled for the result, despite ChatGPT providing the correct link. 
% Upon not finding an appropriate link from Google search, P6 reverted to ChatGPT and clicked the returned link with suspicion to check if it is the correct answer to configure the REST API.
% For example, when P6 prompted ChatGPT for setting up a REST API token, ChatGPT provided the correct link to the relevant GitHub documentation. However, P6 did not immediately trust the link and instead verified it through an independent search:

\begin{small}
\participantquote{I always need to double-check what ChatGPT gives me. It’s not about whether it’s wrong — I just want to be sure.}{P6}    
\end{small}

\noindent This behavior highlights a trust gap for these familiar but not fluent tasks. Developers value the crowd-sourced data and seek supporting evidence, a feature not yet provided by AI assistants alone.
An action pattern reflects an iterative, feedback-driven workflow:
\begin{implicationbox}
    \emph{Reading and Comprehension} → \emph{Crafting Query/Prompt} → \emph{Writing New Code} → \emph{Reading and Comprehension}
\end{implicationbox}

\noindent Developers frequently cycled between reviewing task information, generating AI suggestions, and modifying code based on feedback. From the interviews, we learned that developers often viewed AI suggestions as drafts rather than final solutions:

\begin{small}
\participantquote{I usually take what Copilot gives me and try it out. \textbf{If it works, I keep it; if not, I tweak it or ask ChatGPT to refine it.}}{P7}    
\end{small}

\noindent This back-and-forth process showed that developers are not passively accepting AI-generated code; they remain actively engaged, evaluating and modifying suggestions as needed.

\subsection{Emotional Responses to AI Assistance: Stability, Guilt, and Self-Perception}

As one of the key dimensions of our model, participants highlighted the significance of emotions to their development behavior. 
While AI-assisted participants often experienced fewer emotional fluctuations, they also struggled with feelings of guilt and self-doubt when relying too heavily on AI-generated suggestions. 
This trade-off encapsulates the emotional balance that developers try to navigate between productivity and understanding in AI-assisted software development.
Ultimately, the goal is to grasp the underlying logic of the code. 
\begin{small}
\participantquote{I don’t want to just copy-paste. I need to know what this code actually does.}{P10}    
\end{small}

\subsubsection{The Tension Between Speed and Understanding}

While AI assistance resulted in higher emotional stability and faster completion times, many AI-assisted participants reported guilty feelings. 
The 25-minute time limit nudged participants to prioritize speed over understanding. 
A consequence of prioritizing speed and completion of the tasks was that it left some participants feeling conflicted about the depth of their learning:

\noindent Some participants expressed guilt for relying too much on AI rather than developing their own understanding:

\begin{small}
\participantquote{I finished the REST API integration with ChatGPT's help, \textbf{but I feel guilty — like I didn't learn anything.}}{P12}
\end{small}

\noindent This emotional conflict reflects a deeper tension between efficiency and skill development. 
Previous literature has identified ``improving'' (i.e., continuously building skills) and ``being productive'' (i.e., achieving results faster) as key attributes of great software engineers~\cite{li2015makes}. 
Our study suggests that AI-assisted coding creates a trade-off between these two attributes, generating internal conflict when developers feel they are sacrificing learning for efficiency.

\subsubsection{Impostor Phenomenon and Self-Criticism}
This emotional tension was further compounded by manifestations of the \emph{impostor phenomenon}, which is the belief that one's success is undeserved or due to luck rather than competence. 
Researchers have found \emph{impostor phenomenon} to negatively impact perceived productivity in software engineers~\cite{guenes2024impostor}.

\noindent \textbf{Participants often attributed an issue to their own perceived shortcomings}, such as unclear prompting or technical missteps, rather than questioning the answer provided by an AI assistant, even if the provenance of an issue lies with the tooling. 

\begin{small}
\participantquote{ChatGPT is always right. If the output is wrong, I must’ve messed up the prompt.}{P12}
\end{small}

\noindent This self-criticism carried over into debugging and troubleshooting. When P2 encountered a problem with AI-generated code, their first instinct was to assume that the fault lay with their machine rather than the AI.

\begin{small}
\participantquote{My first thought was, `Is my laptop broken?' I wasted 3 minutes checking my laptop.}{P2}
\end{small}

\noindent Rather than acknowledging limitations that exist in AI assistants, some participants \textbf{inherently assumed faults for any failures.}

\subsection{Show Me, Don’t Tell Me: The Need for Visual Clarity in AI Assistants}
The intention \textit{\textbf{To Obtain Required Knowledge}} was important because this sequence emerged as a notable finding during follow-up interviews. 
Participants reported that this intention was time-consuming, especially when dealing with procedural tasks such as configuring API tokens or looking up the related documentation. 
Although standard AI assistant prompting is straightforward, participants found text-only AI suggestions insufficient for providing in-depth explanations.
For example, when ChatGPT provided written instructions for GitHub API setup, developers like P1 and P4 abandoned the assistant altogether.
They turned to platforms like Stack Overflow or GitHub's official documentation to find step-by-step instructions. 

\begin{small}
\participantquote{ChatGPT told me to `add the token' but where exactly? Some \textbf{screenshots or visual steps would’ve saved me 10 minutes of guessing.}}{P1} 
\end{small}

\noindent This challenge was reflected in a recurring \textbf{action pattern} involving \emph{searching} for additional information when AI suggestions lacked sufficient detail. Although this pattern did not reach our 10\% support threshold in CloSpan, it was repeatedly discussed in interviews. 

\noindent One of the key features missing in current AI assistants is visual guidance. 
When AI-generated text-based instructions were unclear, interviewees frequently abandoned the AI assistants and turned to alternative sources to find more reliable guidance. This led to an action sequence:

% \yinan{so this pattern did not meet 10\%, so we choose to discuss this pattern in discussion section}

\begin{implicationbox}
{\emph{Reading and Comprehension} → \emph{Search} → \emph{Reading and Comprehension}}
\end{implicationbox}

\noindent This pattern highlights a modality mismatch: when a developer's Intention is \textit{\textbf{To Set Up Project Environment}}, the purely text-based modality of current AI assistants fails, forcing a fallback to traditional Search Engines to find diagrams or visual documentation. This pattern reflects developers' need for \textbf{concrete visual guidance} versus abstract textual explanations. P4 expressed greater trust in answers when supported by visual aids:

\begin{small}
\participantquote{I trust answers more when I see some screenshots of the actual settings page.}{P4}
\end{small}

\noindent 
% The problem stemmed from the format of AI-generated suggestions, as text-based guidance often failed to provide the visual clarity.
Similarly, P7 spent a lot of time on the database setup and connection for the Java assignment. 
Despite frequently prompting an AI assistant to debug the database setup and connection errors, these kinds of tasks were more complex than simply copying and pasting a few lines of code. 
The AI assistant provided various instructional sets, but could not visually demonstrate to the participant the correct order of steps to take to achieve the goal.
This forced participants to repeatedly interpret textual suggestions, resulting in high cognitive load and time costs. 
Future work should investigate how to help developers of AI assistants add features that support visual guidance.

%% file: sections/discussion.tex
\section{Discussion and Implications}

\subsection{Situating S-IASE in the Landscape of Developer–AI Interaction Research}
There are significant differences between our proposed S-IASE model and prior work on developer-AI interaction. For example, while Google’s SIA model~\cite{googlesia} describes coding states primarily through intention and action, our model incorporates explicit dimensions of supporting tool and emotion. This addition essentially highlights how affective states (i.e., emotions) intertwine with cognitive and behavioral processes (i.e., actions) during programming, complementing and refining the abstractions used in earlier models such as SIA~\cite{googlesia} and CUPS~\cite{mozannar2024reading}. 
In doing so, our work challenges the view of developer states as a single coding activity, showing instead that intentions, actions, tools, and emotions evolve together as developers interact and collaborate with AI assistants.

\subsection{Implications for Researchers}

\noindent$\bullet$ \textbf{Model for developer-AI programming behavior:} The rapid growth and adoption of AI assistants in software development have placed greater emphasis on understanding how developers use them.
The model proposed in our work is a step toward a deeper understanding of all the intricacies involved in developer behavior when they interact with AI assistants for coding.
Future work could explore some of the behavior patterns uncovered in more depth or incorporate additional dimensions to enhance the model. % so to assist in developing more comprehensive LLMs for software engineering.

\noindent$\bullet$ \textbf{Trusting AI assistants:} Our findings indicate that developers are more willing to use AI assistants when they are highly familiar or completely unfamiliar with a task.
This differs from previous literature on factors impacting developer trust in AI suggestions, such as quality of the suggestion and context~\cite{brown2024identifying}.
Further research should focus on deriving deeper insights about what influences developer trust in AI-generated answers, and more importantly, exploring how AI assistants should be enhanced to improve developer trust.

\noindent$\bullet$ \textbf{Guilt when using AI:} Impostor syndrome has been documented to occur in software engineering and negatively impact perceived productivity~\cite{guenes2024impostor}. 
We found developers self-criticizing during the use of AI assistants for programming and blaming themselves for over-relying on AI. 
Future research should empirically investigate the extent of the impostor phenomenon in AI-assisted development and uncover strategies to mitigate these negative emotions from using AI.

\subsection{Implications for Practitioners}

\noindent$\bullet$ \textbf{Informing industrial AI assistant design:} While we propose the S-IASE model in a research-based study, we expect that this model can be applied in industry as well. 
We know the AI for coding companies continuously leverage user interaction data to refine their models.
Our proposed model may lead the way to systematically analyze user programming behavior and improve the models based on users' real intention, action, and emotion.

\noindent$\bullet$ \textbf{Exploratory software development:} Developers in our study often used AI assistants to conduct exploratory software development, a form of vibe coding phenomenon gaining popularity \cite{vibecoding, sarkar2025vibe}.
AI assistants can generate simple code snippets, allowing developers to focus on the important logic. 
Future research could explore how to incorporate exploratory software development in computing education.
From a practitioner's perspective, software organizations should prepare the proper tools and training to support developers as they follow this software development approach. 

\noindent$\bullet$ \textbf{Toward automated behavior annotation.}
Another promising direction is using our annotated dataset to train models to automatically detect parts of the S-IASE model, specifically the parts about intention, action, and supporting tools, and then prompt participants to reflect on their \textit{Emotion}s.
This semi-automated pipeline could significantly reduce participant workload in future user studies, while enabling large-scale collection of fine-grained developer behavior data.

%% file: sections/2-Related-Work.tex
\subsection{AI Assistant Adoption in Software Engineering}

The rapid adoption of AI assistants is transforming how developers perform software engineering tasks. Programming assistants like GitHub Copilot~\cite{copilot}, GitHub Copilot Chat~\cite{copilot_chat}, 
ChatGPT~\cite{chatgpt}, Codex~\cite{codex}, and Cursor~\cite{cursor} are now widely integrated into software development environments. These AI assistants offer a variety of support from code comprehension, code suggestions, bug fixes, to documentation support.
Prior studies have explored the benefits, and challenges~\cite{prather2023s, liang2024large, nam2024using, li2024ai} of AI assistant adoption in software engineering. 
For example, Liang et al.~\cite{liang2024large} found that developers primarily use AI assistants to reduce keystrokes, accelerate task completion, and recall syntax. They also reported that developers face challenges when AI-generated code fails to meet functional or non-functional requirements which leads to low acceptance rates.

The influence between developers and AI assistants is mutual. Within the loop of software development process, developers not only actively make decisions on top of AI outputs, but also their behavior is subtly impacted by AI assistants in various activities, such as program comprehension and workflow management~\cite{nguyen2024beginning,nam2024using,rajbhoj2024accelerating}. This broader and gradual impact suggests that AI assistant adoption in software engineering extends beyond individual programming tasks to shaping overall software development practices. In this work, we aim to expand this understanding by analyzing how developers' intentions, actions, supporting tools, and emotions interweave with AI assistance in practical software engineering tasks.

\subsection{Developer-AI Interaction in Software Development}

Previous studies have identified specific developer interaction patterns with AI assistants~\cite{treude2025developers, prather2023s, barke2023grounded}. For example, Barke et al.~\cite{barke2023grounded} adopted grounded theory to develop two modes of developer-AI interaction, one is in acceleration mode, the programmer knows what to do next and uses Copilot to get there faster; in exploration mode, the programmer is unsure how to proceed and uses Copilot to explore their options. Similarly, Prather et al.~\cite{prather2023s} also described two common patterns: shepherding (guiding AI suggestions) and drifting (exploring without clear goals).
Amoozadeh et al.~\cite{amoozadeh2024student} not only found that CS1 students showed two similar patterns of ChatGPT use (using as a shortcut or iterative refinement), they also proposed a taxonomy of developer activities with ChatGPT, including \textit{reading}, \textit{prompting}, and \textit{verification}.
However, these studies mainly derive interaction patterns from researchers’ external observations.
In contrast, Mozannar et al.~\cite{mozannar2024reading} directly examined developer–AI interaction from the developer’s perspective by asking participants to self-report their coding states. They developed a CUPS (CodeRec User Programming States) taxonomy with a specific focus on GitHub Copilot, such as \textit{deferring thought for later}. They aim to categorize developer states during code completion with GitHub Copilot.
Beyond physical or observed interaction~\cite{treude2025developers}, \textit{cognition} such as trust perspective~\cite{johnson2023make,sabouri2025trust,wang2024investigating} has also emerged as a key factor in developer-AI interaction.
Choudhuri et al.~\cite{choudhuri2024guides} examined how trust and system quality influenced developers' interactions with AI assistants like GitHub Copilot and ChatGPT. They found that functional value, ease of use, and perceived reliability significantly affected developers’ trust and willingness to adopt AI assistants. 
However, prior studies often treat interaction, behavior, and cognition separately, neglecting their dynamic interplay~\cite{mozannar2024reading}.
Additionally, existing research typically constrains tool usage to specific AI assistants~\cite{amoozadeh2024student, mozannar2024reading, barke2023grounded}, limiting generalizability to real-world contexts.

Unlike prior research, we permitted participants to select any AI assistants they commonly used, capturing developer-AI interaction patterns in their more natural environment. Our model describes  developers' intentions, subsequent actions with supporting tools, and emotions given a certain programming state.
In this way, we aim to create a dynamic feedback loop that accurately reflects real-world software development.

%% file: main.bib
@inproceedings{alomar2024refactor,
  title={How to refactor this code? An exploratory study on developer-ChatGPT refactoring conversations},
  author={AlOmar, Eman Abdullah and Venkatakrishnan, Anushkrishna and Mkaouer, Mohamed Wiem and Newman, Christian and Ouni, Ali},
  booktitle={Proceedings of the 21st International Conference on Mining Software Repositories},
  pages={202--206},
  year={2024},
  doi={10.1145/3643991.3645081}
}

@article{yu2025test,
  title={Test Script Intention Generation for Mobile Application via GUI Image and Code Understanding},
  author={Yu, Shengcheng and Fang, Chunrong and Liu, Jia and Chen, Zhenyu},
  journal={ACM Transactions on Software Engineering and Methodology},
  year={2025},
  publisher={ACM New York, NY},
  doi={10.1145/3722105}
}

@incollection{zimmermann2016card,
  title={Card-sorting: From text to themes},
  author={Zimmermann, Thomas},
  booktitle={Perspectives on data science for software engineering},
  pages={137--141},
  year={2016},
  publisher={Elsevier},
  doi={10.1016/B978-0-12-804206-9.00027-1}
}

@inproceedings{ford2017characterizing,
  title={Characterizing software engineering work with personas based on knowledge worker actions},
  author={Ford, Denae and Zimmermann, Tom and Bird, Christian and Nagappan, Nachiappan},
  booktitle={2017 ACM/IEEE International Symposium on Empirical Software Engineering and Measurement (ESEM)},
  pages={394--403},
  year={2017},
  organization={IEEE},
  doi={10.1109/ESEM.2017.54}
}

@article{harper2012member,
  title={Member checking: Can benefits be gained similar to group therapy},
  author={Harper, Melissa and Cole, Patricia},
  journal={The qualitative report},
  volume={17},
  number={2},
  pages={510--517},
  year={2012},
  doi={10.46743/2160-3715/2012.2139}
}

@article{graziotin2015feelings,
  title={Do feelings matter? On the correlation of affects and the self-assessed productivity in software engineering},
  author={Graziotin, Daniel and Wang, Xiaofeng and Abrahamsson, Pekka},
  journal={Journal of Software: Evolution and Process},
  volume={27},
  number={7},
  pages={467--487},
  year={2015},
  publisher={Wiley Online Library},
  doi={10.1002/smr.1673}
}

@inproceedings{cheng2023multi,
  title={Multi-modal emotion recognition for enhanced requirements engineering: a novel approach},
  author={Cheng, Ben and Arora, Chetan and Liu, Xiao and Hoang, Thuong and Wang, Yi and Grundy, John},
  booktitle={2023 IEEE 31st International Requirements Engineering Conference (RE)},
  pages={299--304},
  year={2023},
  organization={IEEE},
  doi={10.1109/RE57278.2023.00039}
}

@article{madampe2025emoreflex,
  title={EmoReflex: an AI-powered emotion-centric developer insights platform},
  author={Madampe, Kashumi and Grundy, John and Nguyen, Minh and Welstead-Cloud, Ellen and Huynh, Vinh Tuan and Doan, Linh and Lay, William and Hashim, Sayed},
  journal={Automated Software Engineering},
  volume={32},
  number={1},
  pages={22},
  year={2025},
  publisher={Springer},
  doi={10.1007/s10515-025-00488-7}
}

@inproceedings{treude2025developers,
  title={How Developers Interact with AI: A Taxonomy of Human-AI Collaboration in Software Engineering},
  author={Treude, Christoph and Gerosa, Marco A},
  booktitle={In Proceedings of the 2nd ACM International Conference on Al Foundation Models and Software Engineering},
  year={2025},
  doi={10.1109/Forge66646.2025.00033}
}

@inproceedings{nguyen2024beginning,
  title={How beginning programmers and code llms (mis) read each other},
  author={Nguyen, Sydney and Babe, Hannah McLean and Zi, Yangtian and Guha, Arjun and Anderson, Carolyn Jane and Feldman, Molly Q},
  booktitle={Proceedings of the 2024 CHI Conference on Human Factors in Computing Systems},
  pages={1--26},
  year={2024},
  doi={10.1145/3613904.3642706}
}

@inproceedings{rajbhoj2024accelerating,
  title={Accelerating software development using generative ai: Chatgpt case study},
  author={Rajbhoj, Asha and Somase, Akanksha and Kulkarni, Piyush and Kulkarni, Vinay},
  booktitle={Proceedings of the 17th innovations in software engineering conference},
  pages={1--11},
  year={2024},
  doi={10.1145/3641399.3641403}
}

@inproceedings{johnson2023make,
  title={Make your tools sparkle with trust: The PICSE framework for trust in software tools},
  author={Johnson, Brittany and Bird, Christian and Ford, Denae and Forsgren, Nicole and Zimmermann, Thomas},
  booktitle={2023 IEEE/ACM 45th International Conference on Software Engineering: Software Engineering in Practice (ICSE-SEIP)},
  pages={409--419},
  year={2023},
  organization={IEEE},
  doi={10.1109/ICSE-SEIP58684.2023.00043}
}

@inproceedings{sabouri2025trust,
  title={Trust dynamics in AI-assisted development: Definitions, factors, and implications},
  author={Sabouri, Sadra and Eibl, Philipp and Zhou, Xinyi and Ziyadi, Morteza and Medvidovic, Nenad and Lindemann, Lars and Chattopadhyay, Souti},
  booktitle={Proceedings of the 47th IEEE/ACM international conference on software engineering},
  year={2025},
  doi={10.1109/ICSE55347.2025.00199}
}

@inproceedings{choudhuri2024guides,
  title={What guides our choices? Modeling developers' trust and behavioral intentions towards genai},
  author={Choudhuri, Rudrajit and Trinkenreich, Bianca and Pandita, Rahul and Kalliamvakou, Eirini and Steinmacher, Igor and Gerosa, Marco and Sanchez, Christopher and Sarma, Anita},
  booktitle={2025 IEEE/ACM 47th International Conference on Software Engineering (ICSE)},
  pages={1691--1703},
  year={2025},
  doi={10.1109/ICSE55347.2025.00087}
}

@misc{copilot,
  author       = "{GitHub}",
  title        = "{GitHub Copilot}",
  year         = {2023},
  url          = {https://github.com/features/copilot},
}

@misc{copilot_chat,
  author       = "{GitHub}",
  title        = "{GitHub Copilot Chat}",
  year         = {2023},
  url          = {https://github.com/features/copilot},

}

@misc{codex,
  author       = "{OpenAI}",
  title        = "{Codex}",
  year         = {2023},
  url          = {https://openai.com/codex/},

}

@misc{chatgpt,
  author       = "{OpenAI}",
  title        = "{ChatGPT}",
  year         = {2023},
  url          = {https://openai.com/chatgpt},

}

@misc{cursor,
  author       = "{Cursor}",
  title        = "{Cursor}",
  year         = {2023},
  url          = {https://cursor.so},

}

@inproceedings{liang2024large,
  title={A large-scale survey on the usability of ai programming assistants: Successes and challenges},
  author={Liang, Jenny T and Yang, Chenyang and Myers, Brad A},
  booktitle={Proceedings of the 46th IEEE/ACM international conference on software engineering},
  pages={1--13},
  year={2024},
  doi={10.1145/3597503.3608128}
}

@inproceedings{nam2024using,
  title={Using an llm to help with code understanding},
  author={Nam, Daye and Macvean, Andrew and Hellendoorn, Vincent and Vasilescu, Bogdan and Myers, Brad},
  booktitle={Proceedings of the IEEE/ACM 46th International Conference on Software Engineering},
  pages={1--13},
  year={2024},
  doi={10.1145/3597503.3639187}
}

@article{li2024ai,
  title={AI tool use and adoption in software development by individuals and organizations: a grounded theory study},
  author={Li, Ze Shi and Arony, Nowshin Nawar and Awon, Ahmed Musa and Damian, Daniela and Xu, Bowen},
  journal={arXiv preprint arXiv:2406.17325},
  year={2024},
  doi={10.48550/arXiv.2406.17325}

}

@misc{copilot_users,
	title = {Microsoft has over a million paying {Github} {Copilot} users: {CEO} {Nadella}},
	shorttitle = {Microsoft has over a million paying {Github} {Copilot} users},
	url = {https://www.zdnet.com/article/microsoft-has-over-a-million-paying-github-copilot-users-ceo-nadella/},
	language = {en},
	author = {Ray, Tiernan},
        year = {2023},
        month = {10},
        day = {25},
	urldate = {2025-03-01},
	journal = {ZDNET},
}

@article{robinson2024requirements,
  title={Requirements are all you need: The final frontier for end-user software engineering},
  author={Robinson, Diana and Cabrera, Christian and Gordon, Andrew D and Lawrence, Neil D and Mennen, Lars},
  journal={ACM Transactions on Software Engineering and Methodology},
  volume={34},
  number={5},
  pages={1--22},
  year={2025},
  publisher={ACM New York, NY},
  doi={10.1145/3708524}
}

@article{terragni2025future,
  title={The Future of AI-Driven Software Engineering},
  author={Terragni, Valerio and Vella, Annie and Roop, Partha and Blincoe, Kelly},
  journal={ACM Transactions on Software Engineering and Methodology},
  year={2025},
  publisher={ACM New York, NY},
  doi={10.1145/3715003}
}

@INPROCEEDINGS{9653427,
  author={Robles-Aguilar, Alfonso and Ocharán-Hernández, Jorge Octavio and Sánchez-García, Ángel J. and Limón, Xavier},
  booktitle={2021 9th International Conference in Software Engineering Research and Innovation (CONISOFT)}, 
  title={Software Design and Artificial Intelligence: A Systematic Mapping Study}, 
  year={2021},
  volume={},
  number={},
  pages={132-141},
  doi={10.1109/CONISOFT52520.2021.00028}}

@misc{marchezan2024modelbasedmaintenanceevolutiongenai,
      title={Model-based Maintenance and Evolution with GenAI: A Look into the Future}, 
      author={Luciano Marchezan and Wesley K. G. Assunção and Edvin Herac and Alexander Egyed},
      year={2024},
      eprint={2407.07269},
      archivePrefix={arXiv},
      primaryClass={cs.SE},
      url={https://arxiv.org/abs/2407.07269}, 
}

@book{anderson2001taxonomy,
  title={A taxonomy for learning, teaching, and assessing: A revision of Bloom's taxonomy of educational objectives: complete edition},
  author={Anderson, Lorin W and Krathwohl, David R},
  year={2001},
  publisher={Addison Wesley Longman, Inc.}
}

@book{bloom1956taxonomy,
  title={Taxonomy of educational objectives: The classification of educational goals. Handbook 1: Cognitive domain},
  author={Bloom, Benjamin S and Engelhart, Max D and Furst, Edward J and Hill, Walker H and Krathwohl, David R and others},
  year={1956},
  publisher={Longman New York}
}

@INPROCEEDINGS{10449663,
  author={Aleti, Aldeida},
  booktitle={2023 IEEE/ACM International Conference on Software Engineering: Future of Software Engineering (ICSE-FoSE)}, 
  title={Software Testing of Generative AI Systems: Challenges and Opportunities}, 
  year={2023},
  volume={},
  number={},
  pages={4-14},
  doi={10.1109/ICSE-FoSE59343.2023.00009}}

@misc{bidollahkhani2024revolutionizingreliabilityroleai,
      title={Revolutionizing System Reliability: The Role of AI in Predictive Maintenance Strategies}, 
      author={Michael Bidollahkhani and Julian M. Kunkel},
      year={2024},
      eprint={2404.13454},
      archivePrefix={arXiv},
      primaryClass={cs.AI},
      url={https://arxiv.org/abs/2404.13454}, 
}

@misc{chatgpt_users,
	title = {{ChatGPT} now has over 300 million weekly users},
	url = {https://www.theverge.com/2024/12/4/24313097/chatgpt-300-million-weekly-users},
	language = {en-US},
	urldate = {2025-03-01},
	journal = {The Verge},
	author = {Roth, Emma},
	month = dec,
	year = {2024},
}

@inproceedings{amoozadeh2024student,
  title={Student-AI Interaction: A Case Study of CS1 students},
  author={Amoozadeh, Matin and Nam, Daye and Prol, Daniel and Alfageeh, Ali and Prather, James and Hilton, Michael and Srinivasa Ragavan, Sruti and Alipour, Amin},
  booktitle={Proceedings of the 24th Koli Calling International Conference on Computing Education Research},
  pages={1--13},
  year={2024},
  doi={10.1145/3699538.3699567}
}

@misc{torka2024optimizingaiassistedcodegeneration,
      title={Optimizing AI-Assisted Code Generation}, 
      author={Simon Torka and Sahin Albayrak},
      year={2024},
      eprint={2412.10953},
      archivePrefix={arXiv},
      primaryClass={cs.SE},
      url={https://arxiv.org/abs/2412.10953}, 
}

@article{yeticstiren2023evaluating,
  title={Evaluating the code quality of ai-assisted code generation tools: An empirical study on github copilot, amazon codewhisperer, and chatgpt},
  author={Yeti{\c{s}}tiren, Burak and {\"O}zsoy, I{\c{s}}{\i}k and Ayerdem, Miray and T{\"u}z{\"u}n, Eray},
  journal={arXiv preprint arXiv:2304.10778},
  year={2023},
  doi={10.48550/arXiv.2304.10778}
}

@article{hoda2024qualitative,
  title={Qualitative research with socio-technical grounded theory},
  author={Hoda, Rashina},
  journal={Springer},
  year={2024},
  publisher={Springer}
}

@article{hoda2021socio,
  title={Socio-technical grounded theory for software engineering},
  author={Hoda, Rashina},
  journal={IEEE Transactions on Software Engineering},
  volume={48},
  number={10},
  pages={3808--3832},
  year={2021},
  publisher={IEEE},
  doi={10.1109/TSE.2021.3106280}
}

@misc{stackoverflow_survey,
	title = {{AI} {\textbar} 2024 {Stack} {Overflow} {Developer} {Survey}},
	url = {https://survey.stackoverflow.co/2024/ai},
	language = {en},
	urldate = {2025-03-01},
year={2024},
}

@misc{stackoverflow_survey_2025,
	title = {{2025} {Developer} {Survey}},
	url = {https://survey.stackoverflow.co/2025/},
	language = {en},
year = {2025}, 
	urldate = {2025-09-04},
	
}

@misc{stackoverflowtraffic,
  author = {Stack Overflow},
  title = {Insights into Stack Overflow's traffic},
  year = {2023},
  url = {https://stackoverflow.blog/2023/08/08/insights-into-stack-overflows-traffic/},
  note = {Accessed: 03 August 2023}
}

@misc{vibecoding,
  author = {Wikipedia},
  title = {Vibe Coding},
  year = {2025},
  url = {https://en.wikipedia.org/wiki/Vibe_coding},
  note = {Accessed: 2025}
}

@misc{googlesia,
  author = {Google Research},
  title = {Large sequence models for software development activities},
  year = {2023},
  url = {https://research.google/blog/large-sequence-models-for-software-development-activities/},
  note = {Accessed: 2023}
}

@dataset{replication,
  author       = {Wu, Yinan and Li, Ze Shi and Stolee, Kathryn Thomasset and Xu, Bowen},
  title        = {Replication Package},
  month        = April,
  year         = 2026,
  publisher    = {Zenodo},
  doi          = {10.5281/zenodo.19582089},
  url          = {https://doi.org/10.5281/zenodo.19582089}
}

@inproceedings{Kazemitabaar,
author = {Kazemitabaar, Majeed and Chow, Justin and Ma, Carl Ka To and Ericson, Barbara J. and Weintrop, David and Grossman, Tovi},
title = {Studying the effect of AI Code Generators on Supporting Novice Learners in Introductory Programming},
year = {2023},
isbn = {9781450394215},
publisher = {Association for Computing Machinery},
address = {New York, NY, USA},
url = {https://doi.org/10.1145/3544548.3580919},
doi = {10.1145/3544548.3580919},
booktitle = {Proceedings of the 2023 CHI Conference on Human Factors in Computing Systems},
articleno = {455},
numpages = {23},
location = {Hamburg, Germany},
series = {CHI '23}
}

@misc{otterai,
	title = {Otter.ai - {AI} {Meeting} {Note} {Taker} \& {Real}-time {AI} {Transcription}},
	url = {https://otter.ai/},
	language = {en},
	urldate = {2025-03-15},
}

@article{ng2024harnessing,
  title={Harnessing the Potential of Gen-AI Coding Assistants in Public Sector Software Development},
  author={Ng, Kevin KB and Fauzi, Liyana and Leow, Leon and Ng, Jaren},
  journal={arXiv preprint arXiv:2409.17434},
  year={2024},
  doi={10.48550/arXiv.2409.17434}
}

@article{10.1145/3643774,
author = {Murali, Vijayaraghavan and Maddila, Chandra and Ahmad, Imad and Bolin, Michael and Cheng, Daniel and Ghorbani, Negar and Fernandez, Renuka and Nagappan, Nachiappan and Rigby, Peter C.},
title = {AI-Assisted Code Authoring at Scale: Fine-Tuning, Deploying, and Mixed Methods Evaluation},
year = {2024},
issue_date = {July 2024},
publisher = {Association for Computing Machinery},
address = {New York, NY, USA},
volume = {1},
number = {FSE},
url = {https://doi.org/10.1145/3643774},
doi = {10.1145/3643774},
journal = {Proc. ACM Softw. Eng.},
month = jul,
articleno = {48},
numpages = {20}
}

@ARTICLE{10705649,
  author={Carleton, Anita and Falessi, Davide and Zhang, Hongyu and Xia, Xin},
  journal={IEEE Software}, 
  title={Generative AI: Redefining the Future of Software Engineering}, 
  year={2024},
  volume={41},
  number={6},
  pages={34-37},
  doi={10.1109/MS.2024.3441889}}

@inproceedings{mozannar2024reading,
  title={Reading between the lines: Modeling user behavior and costs in AI-assisted programming},
  author={Mozannar, Hussein and Bansal, Gagan and Fourney, Adam and Horvitz, Eric},
  booktitle={Proceedings of the CHI Conference on Human Factors in Computing Systems},
  pages={1--16},
  year={2024},
  doi={10.1145/3613904.3641936}
}

@misc{github,
	title = {{GitHub} {Copilot} · {Your} {AI} pair programmer},
	url = {https://github.com/features/copilot},
	language = {en},
	urldate = {2025-03-14},
	journal = {GitHub},
	year = {2025},
}

@inproceedings{zhang2024students,
  title={Students’ perceptions and preferences of generative artificial intelligence feedback for programming},
  author={Zhang, Zhengdong and Dong, Zihan and Shi, Yang and Price, Thomas and Matsuda, Noboru and Xu, Dongkuan},
  booktitle={Proceedings of the AAAI Conference on Artificial Intelligence},
  volume={38},
  number={21},
  pages={23250--23258},
  year={2024},
  doi={10.1609/aaai.v38i21.30372}
}

@article{prather2023s,
  title={“It’s Weird That it Knows What I Want”: Usability and Interactions with Copilot for Novice Programmers},
  author={Prather, James and Reeves, Brent N and Denny, Paul and Becker, Brett A and Leinonen, Juho and Luxton-Reilly, Andrew and Powell, Garrett and Finnie-Ansley, James and Santos, Eddie Antonio},
  journal={ACM Transactions on Computer-Human Interaction},
  volume={31},
  number={1},
  pages={1--31},
  year={2023},
  publisher={ACM New York, NY},
  doi={10.1145/3617367}
}

@article{russell1980circumplex,
  title={A circumplex model of affect.},
  author={Russell, James A},
  journal={Journal of personality and social psychology},
  volume={39},
  number={6},
  pages={1161},
  year={1980},
  publisher={American Psychological Association},
  doi={10.1037/h0077714}
}

@incollection{gardella2024performance,
  title={Performance, Workload, Emotion, and Self-Efficacy of Novice Programmers Using AI Code Generation},
  author={Gardella, Nicholas and Pettit, Raymond and Riggs, Sara L},
  booktitle={Proceedings of the 2024 on Innovation and Technology in Computer Science Education V. 1},
  pages={290--296},
  year={2024},
  doi={10.1145/3649217.3653615}
}

@inproceedings{wang2024investigating,
  title={Investigating and designing for trust in ai-powered code generation tools},
  author={Wang, Ruotong and Cheng, Ruijia and Ford, Denae and Zimmermann, Thomas},
  booktitle={The 2024 ACM Conference on Fairness, Accountability, and Transparency},
  pages={1475--1493},
  year={2024},
  doi={10.1145/3630106.3658984}
}

@inproceedings{brown2024identifying,
  title={Identifying the factors that influence trust in AI code completion},
  author={Brown, Adam and D'Angelo, Sarah and Murillo, Ambar and Jaspan, Ciera and Green, Collin},
  booktitle={Proceedings of the 1st ACM International Conference on AI-Powered Software},
  pages={1--9},
  year={2024},
  doi={10.1145/3664646.3664757}
}

@article{barke2023grounded,
  title={Grounded copilot: How programmers interact with code-generating models},
  author={Barke, Shraddha and James, Michael B and Polikarpova, Nadia},
  journal={Proceedings of the ACM on Programming Languages},
  volume={7},
  number={OOPSLA1},
  pages={85--111},
  year={2023},
  doi={10.1145/3586030},
  publisher={ACM New York, NY, USA}
}

@inproceedings{li2015makes,
  title={What makes a great software engineer?},
  author={Li, Paul Luo and Ko, Amy J and Zhu, Jiamin},
  booktitle={2015 IEEE/ACM 37th IEEE International Conference on Software Engineering},
  volume={1},
  pages={700--710},
  year={2015},
  organization={IEEE},
  doi={10.1109/ICSE.2015.335}
}

@inproceedings{guenes2024impostor,
  title={Impostor phenomenon in software engineers},
  author={Guenes, Paloma and Tomaz, Rafael and Kalinowski, Marcos and Baldassarre, Maria Teresa and Storey, Margaret-Anne},
  booktitle={Proceedings of the 46th International Conference on Software Engineering: Software Engineering in Society},
  pages={96--106},
  year={2024},
  doi={10.1145/3639475.3640114}
}

@inproceedings{girardi2020recognizing,
  title={Recognizing developers' emotions while programming},
  author={Girardi, Daniela and Novielli, Nicole and Fucci, Davide and Lanubile, Filippo},
  booktitle={Proceedings of the ACM/IEEE 42nd international conference on software engineering},
  pages={666--677},
  year={2020},
  doi={10.1145/3377811.3380374}
}

@inproceedings{horvath2022using,
  title={Using annotations for sensemaking about code},
  author={Horvath, Amber and Myers, Brad and Macvean, Andrew and Rahman, Imtiaz},
  booktitle={Proceedings of the 35th Annual ACM Symposium on User Interface Software and Technology},
  pages={1--16},
  year={2022},
  doi={10.1145/3526113.3545667}
}

@article{sarkar2025vibe,
  title={Vibe coding: programming through conversation with artificial intelligence},
  author={Sarkar, Advait and Drosos, Ian},
  journal={arXiv preprint arXiv:2506.23253},
  year={2025},
  doi={10.48550/arXiv.2506.23253}
}

@inproceedings{kam2025professional,
  title={What do professional software developers need to know to succeed in an age of Artificial Intelligence?},
  author={Kam, Matthew and Miller, Cody and Wang, Miaoxin and Tidwell, Abey and Lee, Irene A and Malyn-Smith, Joyce and Perret, Beatriz and Tiwari, Vikram and Kenitzer, Joshua and Macvean, Andrew and others},
  booktitle={Proceedings of the 33rd ACM International Conference on the Foundations of Software Engineering},
  pages={947--958},
  year={2025},
  doi={10.1145/3696630.3727251}
}

@inproceedings{zakharov2025ai,
  title={AI in Software Engineering: Perceived Roles and Their Impact on Adoption},
  author={Zakharov, Ilya and Koshchenko, Ekaterina and Sergeyuk, Agnia},
  booktitle={Proceedings of the 33rd ACM International Conference on the Foundations of Software Engineering},
  pages={1305--1309},
  year={2025},
  doi={10.1145/3696630.3730563}
}

@article{chen2021evaluating,
  title={Evaluating large language models trained on code},
  author={Chen, Mark and Tworek, Jerry and Jun, Heewoo and Yuan, Qiming and Pinto, Henrique Ponde De Oliveira and Kaplan, Jared and Edwards, Harri and Burda, Yuri and Joseph, Nicholas and Brockman, Greg and others},
  journal={arXiv preprint arXiv:2107.03374},
  year={2021},
  doi={10.48550/arXiv.2107.03374}
}

@inproceedings{yan2003clospan,
  title={CloSpan: Mining: Closed sequential patterns in large datasets},
  author={Yan, Xifeng and Han, Jiawei and Afshar, Ramin},
  booktitle={Proceedings of the 2003 SIAM international conference on data mining},
  pages={166--177},
  year={2003},
  organization={SIAM},
  doi={10.1137/1.9781611972733.15}
}

@inproceedings{hart2006nasa,
  title={NASA-task load index (NASA-TLX); 20 years later},
  author={Hart, Sandra G},
  booktitle={Proceedings of the human factors and ergonomics society annual meeting},
  volume={50},
  number={9},
  pages={904--908},
  year={2006},
  organization={Sage publications Sage CA: Los Angeles, CA},
  doi={10.1177/154193120605000909}
}

@article{damevski2016mining,
  title={Mining sequences of developer interactions in visual studio for usage smells},
  author={Damevski, Kostadin and Shepherd, David C and Schneider, Johannes and Pollock, Lori},
  journal={IEEE Transactions on Software Engineering},
  volume={43},
  number={4},
  pages={359--371},
  year={2016},
  publisher={IEEE},
  doi={10.1109/TSE.2016.2592905}
}

@inproceedings{xie2006mapo,
  title={MAPO: Mining API usages from open source repositories},
  author={Xie, Tao and Pei, Jian},
  booktitle={Proceedings of the 2006 international workshop on Mining software repositories},
  pages={54--57},
  year={2006},
  doi={10.1145/1137983.1137997}
}

@book{cohen2013applied,
  title={Applied multiple regression/correlation analysis for the behavioral sciences},
  author={Cohen, Jacob and Cohen, Patricia and West, Stephen G and Aiken, Leona S},
  year={2013},
  publisher={Routledge}
}

@article{long2000using,
  title={Using heteroscedasticity consistent standard errors in the linear regression model},
  author={Long, J Scott and Ervin, Laurie H},
  journal={The American Statistician},
  volume={54},
  number={3},
  pages={217--224},
  year={2000},
  publisher={Taylor \& Francis},
  doi={10.1080/00031305.2000.10474549}
}

@article{benjamini1995controlling,
  title={Controlling the false discovery rate: a practical and powerful approach to multiple testing},
  author={Benjamini, Yoav and Hochberg, Yosef},
  journal={Journal of the Royal statistical society: series B (Methodological)},
  volume={57},
  number={1},
  pages={289--300},
  year={1995},
  publisher={Wiley Online Library},
  doi={10.1111/j.2517-6161.1995.tb02031.x}
}

@article{laird1982random,
  title={Random-effects models for longitudinal data},
  author={Laird, Nan M and Ware, James H},
  journal={Biometrics},
  pages={963--974},
  year={1982},
  publisher={JSTOR},
  doi={10.2307/2529876}
}

@article{liang1986longitudinal,
  title={Longitudinal data analysis using generalized linear models},
  author={Liang, Kung-Yee and Zeger, Scott L},
  journal={Biometrika},
  volume={73},
  number={1},
  pages={13--22},
  year={1986},
  publisher={Oxford University Press},
  doi={10.1093/biomet/73.1.13} 
}
